\documentclass[10pt,conference]{IEEEtran}
\IEEEoverridecommandlockouts
\usepackage[linesnumbered,algoruled,boxed,lined]{algorithm2e}
\usepackage{algpseudocode}
\usepackage{makecell}
\usepackage{amsmath}
\usepackage[numbers]{natbib}
\usepackage{enumitem} 
\usepackage{multirow}
\usepackage{booktabs}
\usepackage{hyperref}
\usepackage{diagbox}
\usepackage{amssymb}
\hypersetup{
    colorlinks=true,
    linkcolor=blue,
    filecolor=magenta,
    urlcolor=magenta,
}
\usepackage[all]{hypcap}
\usepackage{booktabs}
\usepackage{url}
\usepackage{xcolor}
\usepackage{tikz}
\usepackage{diagbox}
\usepackage{listings}
\usepackage{comment} 
\usepackage{graphicx}
\usepackage{subcaption}
\usepackage[framemethod=TikZ]{mdframed}
\usepackage{tcolorbox}
\usepackage{amsthm}

\theoremstyle{definition}

\tcbset{
    colback=gray!10, 
    colframe=gray, 
    coltext=black, 
    boxsep=5pt, 
    arc=0mm, 
    top=0mm, bottom=0mm, 
    left=1mm, right=1mm, 
    boxrule=0.5mm, 
    toptitle=1mm, bottomtitle=1mm, 
    titlerule=0mm, 
}
\definecolor{deepgreen}{rgb}{0.0, 0.4, 0.0}

\definecolor{codegreen}{rgb}{0,0.6,0}
\definecolor{codegray}{rgb}{0.5,0.5,0.5}
\definecolor{codepurple}{rgb}{0.58,0,0.82}
\definecolor{backcolour}{rgb}{0.95,0.95,0.92}

\lstdefinestyle{mystyle}{
  backgroundcolor=\color{backcolour}, commentstyle=\color{codegreen},
  keywordstyle=\color{magenta},
  numberstyle=\tiny\color{codegray},
  stringstyle=\color{codepurple},
  basicstyle=\ttfamily\footnotesize,
  breakatwhitespace=false,         
  breaklines=true,                 
  captionpos=b,                    
  keepspaces=true,                 
  numbers=left,                    
  numbersep=5pt,                  
  showspaces=false,                
  showstringspaces=false,
  showtabs=false,                  
  tabsize=2
}
\usepackage{xurl}

\definecolor{mycolor}{RGB}{194, 214, 236}

\newcounter{finding}

\newcounter{result}

\makeatletter
\g@addto@macro{\@algocf@init}{\SetKwInOut{Parameter}{Parameters}}
\makeatother

\newcommand*\circled[1]{\tikz[baseline=(char.base)]{
            \node[shape=circle, draw, inner sep=0.2pt] (char) {\textcolor{black}{#1}};}}

\newcommand{\tech}{\mbox{\textsc{RuleLLM}}}   
 
\begin{document}

\begin{sloppypar}
\title{Automatically Generating Rules of Malicious Software Packages via Large Language Model}

\author{
    \IEEEauthorblockN{XiangRui Zhang\IEEEauthorrefmark{1}, XueJie Du\IEEEauthorrefmark{1}, HaoYu Chen\IEEEauthorrefmark{1},
    Yongzhong He\IEEEauthorrefmark{1}, Wenjia Niu\IEEEauthorrefmark{1},   Qiang Li\IEEEauthorrefmark{1}\IEEEauthorrefmark{2}\thanks{\IEEEauthorrefmark{2}Qiang Li is the corresponding author; email: liqiang@bjtu.edu.cn}}
\IEEEauthorblockA{\IEEEauthorrefmark{1}\textit{School of Cyberspace Security, Beijing Jiaotong University, China} }
}

\maketitle
\thispagestyle{plain}  

\begin{abstract}

Today's security tools predominantly rely on predefined rules crafted by experts, making them poorly adapted to the emergence of software supply chain attacks.
To tackle this limitation, we propose a novel tool, {\tech}, which leverages large language models (LLMs) to automate rule generation for OSS ecosystems.
{\tech} extracts metadata and code snippets from malware as its input, producing YARA and Semgrep rules that can be directly deployed in software development. 
Specifically, the rule generation task involves three subtasks: crafting rules, refining rules, and aligning rules. 
To validate {\tech}'s effectiveness, we implemented a prototype system and conducted experiments on the dataset of 1,633 malicious packages. 
The results are promising—{\tech} generated 763 rules (452 YARA and 311 Semgrep) with a precision of 85.2\% and a recall of 91.8\%, outperforming state-of-the-art (SOTA) tools and scored-based approaches. 
We further analyzed generated rules and proposed a rule taxonomy: 11 categories and 38 subcategories.

\end{abstract}

\section{Introduction}

Open source software (OSS) has become a cornerstone of software development, enabling developers to build applications efficiently by reusing codebases, libraries, and tools.
Meanwhile, OSS ecosystems are suffering new security challenges, including malware infiltration, supply chain attacks, and vulnerabilities in third-party dependencies.
Attackers can exploit vulnerabilities in popular packages, distribute malware, or insert malicious code into legitimate software. 
Recently, the number of security incidents in OSS ecosystems has increased significantly, with an annual growth rate of 742\% in the past five years~\cite{oss_report, ssc-event, pashchenko2020preliminary}, e.g., 0 incidents in 2018, 249 incidents in 2019, and 690,211 incidents in 2023.
For instance, LOG4J vulnerabilities~\cite{log4j} brought great risks to software systems and third-party packages relying on this library.
This attack propagated across downstream projects, potentially compromising thousands of applications, and underscoring the systemic risks inherent in the OSS ecosystem.

To address these risks, developers and security analysts rely on software detection tools to discover underlying risks in OSS ecosystems.
Most software detection tools, such as SemGrep~\cite{semgrep}, YARA~\cite{naik2020evaluating}, and AppInspector~\cite{appInspector}, predominantly rely on malware signatures and predefined rules.
Those tools are designed for pattern-matching malware and analyzing suspicious files based on textual or binary features. 
In particular, predefined rules leverage built-in specific patterns, strings, and features to identify underlying threats. 
However, crafting detection rules requires manual effort and domain knowledge, which is not scalable for large and diverse software packages in OSS ecosystems.
Moreover, as the number of malware grows, manual written rules struggle to keep pace with emerging threats, lacking adaptability in software supply chain (SSC) attacks.

In this work, we introduce large language models (LLMs) to automate rule generation, serving as a supplement to existing security tools.
Nowadays, the LLM offers an innovative approach to enhance security detection and analysis capabilities, including vulnerability identification~\cite{ullah2023llms} and repair~\cite{pearce2023examining, wang2024repository}, static analysis~\cite{li2023hitchhiker, shang2024far}, and reverse engineering~\cite{pearce2022pop}. 
For example, Li et al.~\cite{li2023hitchhiker} investigated LLMs' capabilities in static analysis for automatically repairing zero-shot vulnerabilities, and Wang et al.~\cite{wang2024repository} explored project-level vulnerability detection via LLMs (e.g., ChatGPT or CodeLlama).
In contrast, we propose leveraging LLMs to enhance security tools' capabilities by automating the creation of rules that are tailored to detect both known and emerging threats. 

However, directly utilizing LLMs for rule generation poses three technical challenges.
First, many malicious packages share similar code bases, leading to redundancy and inaccuracies in the generated rules. 
Second, LLMs struggle to process the extensive source code of many malicious packages, which may exceed their input limitations. 
Third, due to the inherent generation characteristics of LLMs, such as hallucinations, they may produce rules that are unsuitable for direct use in development environments. Additionally, some malware employs obfuscation techniques to conceal its intent, further complicating detection.




To address those challenges, we propose a novel tool, {\tech}, designed to automatically generate YARA \& Semgrep rules for detecting risks in OSS ecosystems.
Rather than relying solely on LLM, {\tech} decomposes rule generation into different subtasks: crafting, refining, and aligning rules. 
In the crafting stage, {\tech} extracts independent code blocks from malware and creates coarse-grained rules based on these blocks.  
In the refining stage, redundant or ineffective rules are merged, ensuring higher specificity and accuracy.
To mitigate the risk of hallucinations of LLMs, {\tech} integrates a specialized LLM-based agent to refine rules.
The agent’s feedback loop allows for dynamic refinement of the generated rules.
The outputs of {\tech} include YARA and Semgrep rules that are ready for direct deployment in security workflows.

To validate the effectiveness of {\tech}, we implemented a prototype system based on several open-source libraries. 
Our dataset comprised 3,200 malware packages collected from GuardDog~\cite{guarddog}, which were reduced to 1,633 unique packages after deduplication. 
We also selected 500 of the most commonly used legitimate packages from \cite{pkg-popular}. 
We benchmarked {\tech} against multiple baselines, including state-of-the-art (SOTA) tools, scored-based approaches, and several diverse LLMs (GPT-3.5, GPT-4o, Llama, Claude). 
Experimental results demonstrate that {\tech} generated 763 rules (452 YARA and 311 Semgrep) with a precision of 85.2\% and a recall of 91.8\% in identifying malicious packages, significantly outperforming the baseline tools.
Notably, the generated rules are fully compatible with existing systems and can be directly deployed to scan software packages without errors. 
Furthermore, we conducted a systematic analysis of the generated rules and proposed a comprehensive rule taxonomy. 
The taxonomy consists of 11 categories and 38 subcategories, offering insights into the characteristics and applications of the rules.


In short, our contributions are as follows:
\begin{itemize}
    \item 
    We have proposed {\tech}, a novel tool~\cite{rule-code} that automatically generates YARA and Semgrep rules for OSS ecosystems.
    \item 
    We have generated 452 YARA and 311 Semgrep rules that are well-formatted and can be directly deployed in existing tools.
    Experimental results demonstrated that {\tech} outperforms SOTA tools.
    \item We have released {\tech}'s tool and 763 compatible rules to the research community, as a supplement to security detection tools.
\end{itemize}

\noindent
\textbf{Roadmap.}
The rest of the paper is organized as follows. 
Section~\ref{sec:back} illustrates the background of LLM and rule-based detection tools.
Section~\ref{sec:design} and \ref{sec:gen} present the design of {\tech}, including the malware information extraction and rule generation.
Section~\ref{sec:exp} evaluates {\tech} on real-world malicious packages and compares it with existing tools. 
Section~\ref{sec:discuss} discuss the limitations of {\tech} and its generated rules.
Section~\ref{sec:related} discusses related work, and Section~\ref{sec:conclusion} concludes the paper.

\section{Background}
\label{sec:back}

In this section, we provide background about LLMs, YARA/Semgrep rules, and the technical challenges.

\subsection{Large Language Model}

Large Language Models (LLMs)~\cite{openai, open-codex, claude, gemini} acts as a generation model that outputs a sequence of tokens to complete given input sequences. 
For instance, an LLM can generate a sequence of prediction tokens (an answer) in response to a sequence of input tokens (a question).  
Tokens refer to common characters with a unique numeric identifier, up to a pre-training corpus.
The input to an LLM is typically structured as a prompt, which defines the sequence of input tokens.  
Different prompts (inputs) can lead to diverse capacities of LLM and determine its effectiveness across various tasks.  
Hence, researchers and developers use prompt engineering to design robust and effective prompts to improve the capacity of LLMs on a wide range of complex tasks.

Here, we illustrate several techniques in prompt engineering for LLMs.
(1) \textit{Task decomposition} is to break down a larger task into the large, complex task into smaller, manageable subtasks. 
Then, LLMs execute each subtask sequentially to ensure the completion of the entire task.
(2) \textit{Chain-of-thought} introduces intermediate reasoning steps to execute a task, enabling LLMs to ``think step by step''~\cite{wei2022chain}.  
(3) \textit{Tree-of-thought} generates multiple thoughts at each step, forming a decision tree.
Then, LLMs use search algorithms, such as breadth-first search (BFS) or depth-first search (DFS), to navigate and execute the task~\cite{yao2023tree}.  
(4) \textit{Self-reflection} is to use LLMs to evaluate their own prior outputs to provide feedback and improve subsequent results.
For instance, previous works~\cite{shinn2023reflexion, liu2023languages} leverage human feedback, such as error corrections, to fine-tune LLMs for enhanced performance.  
In short, prompt engineering is about obtaining the ``best'' results of LLMs when we don’t retrain the model.

In this paper, we propose leveraging prompt engineering techniques to enable LLMs to automatically generate rules for security detection tools.


\subsection{YARA \& Semgrep Rule}

Today's security tools use rule-based detection manner to discover potential vulnerabilities, malware, and security threats. 
Tools leverage these rules to scan OSS packages and to find suspicious entities, such as specific file structures, suspicious functions, unusual network activity, or indicators of insecure configurations.

\begin{table}[!t]
    \begin{tabular}{p{8.0cm}}
    \toprule
      \includegraphics[width=\linewidth]{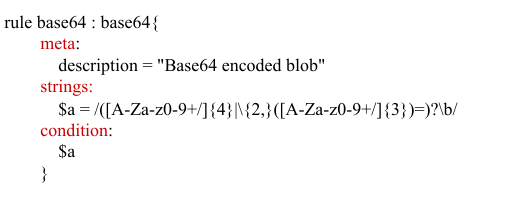} \\ \bottomrule
      \includegraphics[width=\linewidth]{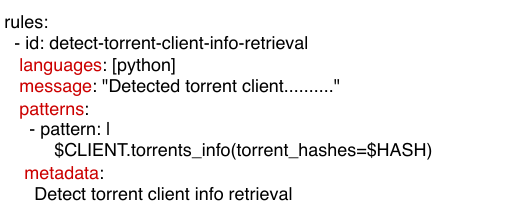} \\ 
    \bottomrule
    \end{tabular}
    \caption{The upper part is a YARA rule, and the below part is a Semgrep rule.} 
    \label{fig:example} 
\end{table}

YARA \& Semgrep rules are two widely used rule formats in cybersecurity, both employing pattern-matching techniques to detect malicious behaviors and threats.  
A YARA rule is a structured, text-based format with a `*.yar' file extension.
It typically consists of four main components: a rule name, a `meta' section, a `Strings' section, and a `Condition' section.  
In contrast, a Semgrep rule is written in YAML format with a `*.yaml' file extension and comprises four key components: a `metadata' section, a `pattern' section, a `languages' section, and a `message' section.  
Table~\ref{fig:example} illustrates two examples of YARA and Semgrep rules.  
The YARA rule detects base64-encoded blobs, while the Semgrep rule identifies torrent information retrieval.  
Both rule formats are designed for clarity and efficiency, offering robust mechanisms for integrating security checks into development workflows.

However, rule-based detection approaches face several limitations.
First, writing a YARA\&Semgrep rule relies on expert knowledge and manual effort, making it challenging to keep up with the rapidly evolving threat landscape. 
Second, YARA\&Semgrep rules are not specifically designed for OSS ecosystems. 
YARA rules focus on identifying patterns in files and binaries, whereas Semgrep rules analyze structured source code. 
Our investigation shows 4,574 YARA and 2,841 Semgrep rules, with most related to email, cloud, mobile, and APT attacks. 
Only 380 rules (46 YARA and 334 Semgrep) are related to OSS packages. 
Third, the growing prevalence of malware and supply chain attacks in OSS ecosystems has made it increasingly challenging to develop rules that comprehensively address all possible risks.

\begin{figure}[!t]  
    \centering
    \includegraphics[width=.85\linewidth]{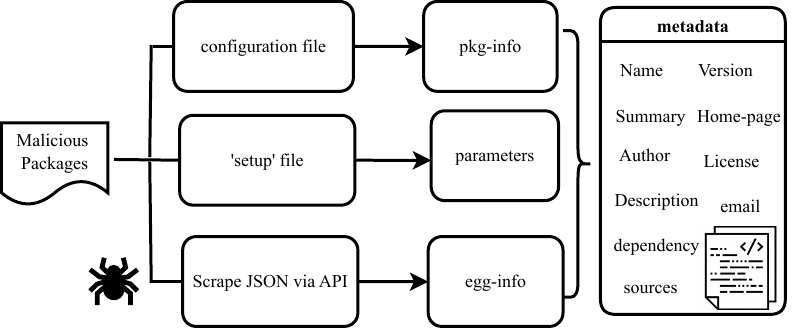}
    \caption{Extracting the metadata of the software package.}
    \label{fig:metadata}
\end{figure}

\subsection{Technical Challenges}

Directly leveraging LLM to generate rules presents three key technical challenges, outlined as follows:  
\begin{itemize}
    \item Malicious packages frequently exhibit diverse behaviors, including privilege escalation, information leakage, code obfuscation, and remote network operations, further complicating accurate rule creation. Additionally, attackers may use anti-analysis techniques to conceal malicious behavior, making it difficult for LLMs to capture features.
    \item The LLM has a context length limitation that determines the maximum volume of information of LLMs' prompts. Many malicious packages have several source code files with token counts that exceed the LLM's context length, resulting in incomplete analyses.
    \item Due to the strict syntax and structural requirements of the rules, LLM may involve errors and hallucinations when generating YARA and Semgrep rules.  
\end{itemize}


To address these technical challenges, we propose {\tech}, a tool that automatically generates YARA and Semgrep rules without any manual effort.  
The generated rules can be directly utilized in existing security tools.


\section{Malware Knowledge Extraction}
\label{sec:design}	

In this section, we present the methodology that automatically extracts metadata and code snippets from a malicious package.

\subsection{Package Metadata}
\label{sec:sub:meta}

Package metadata is information that package authors maintain within the project. 
Specifically, metadata lists dependencies, URLs, versions, package names, and descriptions, which are useful for finding and installing packages from OSS ecosystems.
There are 3 manners to extract its metadata. 
Figure~\ref{fig:metadata} depicts the process of the metadata extraction, including the `egg-info', `package-info', and `setup' files. 
(1) The `pkg-info' file is similar to the project's configuration file, which describes the package's installation and use.
(2) The `setup' file defines the metadata and configuration for a package. 
This file is essential for making the package installable via the OSS ecosystem.
(3) The `egg-info' contains the descriptive information of the package on the OSS ecosystem, which provides an API endpoint for retrieving package metadata.
For instance, an API endpoint from the NPM ecosystem is shown as https://registry.npmjs.org/\{package\_name\}, and its response is a JSON file for the package metadata. 
Given a malicious package, we extract its metadata in those ways.
Note that we use the package metadata as the LLM's input rather than the original package. 

\begin{figure}[!t]  
    \centering
    \includegraphics[width=.85\linewidth]{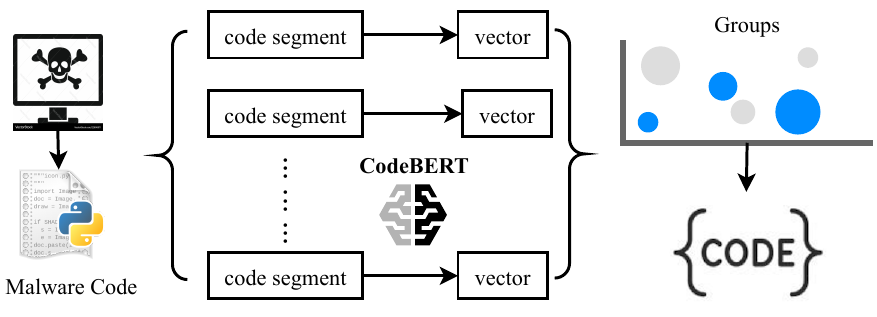}
    \caption{Extracting the package code.}
    \label{fig:code}
\end{figure}

The rationale behind this is that the metadata of the malicious package is different from the benign package. 
First, malicious packages may have minimal, fake, or no author information, but benign packages usually have valid author names, emails, and sometimes links to their profiles or organizations.
Second, malicious packages often have excessive or unusual dependencies, sometimes including outdated or obscure packages that aren't typically used in benign software.
Third, malicious packages often use typo-squatting or lookalike names (e.g., `reqests' instead of `requests'). Descriptions may be vague or directly copied from the legitimate package to mislead users. 
In contrast, benign packages usually have clear, unique names and detailed descriptions.
Hence, rules may be associated with the package metadata, and {\tech} generates the detection rules based on malicious packages' metadata.



\begin{figure*}[!t]  
    \centering
    \includegraphics[width=.86\linewidth]{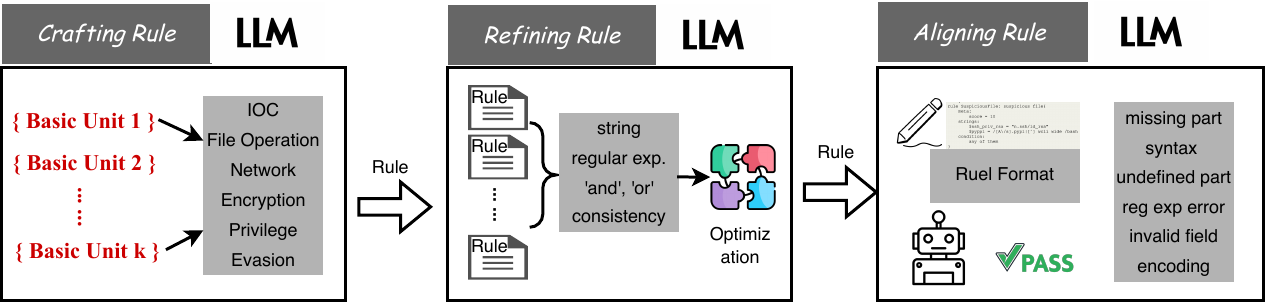}
    \caption{{\tech}: the architecture of the LLM-based rule generation: (1) crafting basic-unit rules, (2) refining rules, and (3) aligning rules.}
    \label{fig:arch:llm}
\end{figure*}

\subsection{Malicious Code Snippets} 
\label{sec:sub:code}

Malicious packages often contain code specifically crafted to perform unwanted or harmful actions on a system.
For example, an unusual `setup.py' or post-install command that executes unwanted code during package installation. 
We propose to extract distinguished code snippets from the malicious package to represent the malicious behavior, as shown in Figure~\ref{fig:code}. 


\textbf{Unpacking}.
To obtain malicious code, we first need to download the package and unpack it to a folder.
Unpacking refers to extracting a package's contents from a compressed or archived format into usable files and folders. 
This operation involves various tasks, such as decompression, file extraction, and directory creation. 
After unpacking a software package, we obtain the source code files, e.g., the file in the PyPI ecosystem uses the extension `.py', while that in the NPM ecosystem uses the extension `.js'.

\textbf{Vectorization}. 
For the source code, we leverage embedding technical to convert source code into numerical vectors.
The embedding represents the content and characteristics of source code in a real-valued continuous vector space.
We use lexical analysis techniques to process the source code: splitting code segments, numerical representation, and concatenation.
(1) Splitting. 
The source code $S_{code}$ is divided into code segments.
For example, $S_{code} = \{code^1, code^2,..., code^n\}$, where $code^i$ presents a code segment, and $n$ is the number of code segments. 
Note each code segment has a fixed length, denoted as a threshold.  
(2) Numerical representation. 
The pre-trained model converts each code segment into numerical representations, which contain information about the structure and semantics of the code.
The formula is $v^i = f(code^i)$, where $v^i$ presents a vector of a code segment. 
(3) Concatenation. 
We concatenate all numerical representations into a single vector, where $V_{code} = \{v^1, v^2,..., v^n\}$ represents the source code $S_{code}$.
Specifically, we use 512 as the length threshold to split the source code. 
The pre-trained model is CodeBert~\cite{feng2020codebert}, and we use the NumPy~\cite{numpy} to join all numerical arrays into one vector.

\textbf{Group}. 
We utilize a clustering algorithm to group similar malware code snippets into the same cluster. 
The similarity between two code snippets is measured using the Euclidean distance in vector space.  
For implementation, we employ the Scikit-learn library~\cite{scikit} to apply the K-Means algorithm.  
In the initialization phase, we set the random seed for the centroids to 42 and define the maximum number of iterations for the K-Means algorithm as 500.  
Clusters with an intra-similarity below 0.85 are discarded, as they lack sufficient homogeneity.  
Conversely, clusters with an intra-similarity of 0.85 or higher are retained.  
After clustering, the similarity among code snippets within a cluster is high, whereas the similarity between different clusters is minimal.  
{\tech} generates rules by analyzing and synthesizing code snippets within each retained cluster.









\section{Rule Generation}
\label{sec:gen}

Our target is to generate YARA \& Semgrep rules that can be seamlessly integrated into existing security tools and effectively detect threats without errors.  
Figure~\ref{fig:arch:llm} presents the architecture of {\tech}, which utilizes an LLM to automate rule generation.  
The rule generation is divided into three subtasks: crafting rules, refining rules, and aligning rules.  
(1) This component involves randomly selecting several basic units from code or metadata groups. 
In terms of those basic units, the LLM crafts rules to cover possible features and patterns associated with malicious packages. 
(2) The LLM audits all coarse-grained rules from the first subtask.
Then, the LLM merges rules into a scalable and effective rule.
(3) An LLM-based agent determines whether the generated rule can pass the compilation process.
The agent is equipped with two creation functions that compile rules.
If a rule successfully compiles, we output the finalized rule.  
If a rule fails, the compiler outputs error messages, and the agent leverages error messages as the guideline to fix rules.
In the following, we elaborate on the details of those components in {\tech}.

\subsection{Crafting Rules} 
\label{sec:task1}

\begin{table}[!t] 
    \caption{Identifying malicious behaviors from basic units.  }
    \label{tab:unit}
    \centering
    \begin{tabular}{  l r   }
    \toprule
    Metadata                    &  Description          \\
    \toprule
    \multirow{1}*{Empty information}  &   \makecell[r]{an empty description}   \\
    Release zero            &  \makecell[r]{ a version like 0.0 or 0.0.0}    \\  
    Typosquatting	&   similar name to a popular package \\
    Dependencies  & malicious dependency libraries \\
    \toprule   
    Code                    &  Description          \\ 
    \toprule  
     IOC &  \makecell[r]{Identify compromised indicators or \\ technical behaviors in code segments} \\
    File &  \makecell[r]{Find file operations like `open()`, `write()`,\\ `remove()` or suspicious file paths} \\
    Network  &    \makecell[r]{ Detect API calls or requests for \\C2 server connections or data exfiltration } \\
    Encryption            &   \makecell[r]{Identify the use of encryption algorithms \\ like `AES`, `RSA`, or base64 encoding }   \\
    Privilege  &\makecell[r]{Identify privilege escalation like\\ `setuid()', `setgid()' or `CreateProcess()'} \\
    \makecell[l]{Anti-debug\\/Anti-analysis}& \makecell[r]{ Detect functions for debuggers, sandbox \\ environments or VM detection techniques}\\

    \toprule    
    \end{tabular}
\end{table}

\textbf{Basic Unit}  
We divide the metadata and code of malware into basic units, which serve as the foundation for creating coarse-grained rules.

\textit{(1) Metadata} is extracted as a JSON format as described in Section~\ref{sec:sub:meta}.
The entire metadata of a package is treated as a base unit.
Several metadata attributes may indicate malicious behaviors, including malicious dependency libraries, empty descriptions, zero versions, and typosquatting. 
Table~\ref{tab:unit} lists four audit categories for package metadata.
\circled{1} Empty information: the package has an empty description. 
\circled{2} Release zero: the package has a version like 0.0 or 0.0.0. 
\circled{3} Typosquatting: the package has a similar name to a popular package. 
\circled{4} Dependencies: the package has malicious dependency libraries. 
We only focus on the suspicious parts of the metadata.

\textit{(2) Code} is organized into different groups as described in Section~\ref{sec:sub:code}, where snippets within the same group exhibit high similarity.
However, due to the LLM's input length limitation and the complexity of the code, entire code snippets cannot be directly processed. 
Each code snippet is divided into multiple basic units to address this.
A basic unit represents a code block: a module, a function body, and a class definition. 
Based on the definition provided in the Python documentation~\footnote{https://docs.python.org/3/reference/executionmodel.html}, our extraction process follows these steps: \circled{1} Use regex to identify whether the code begins with a specific string (e.g., 'def ', 'class ', 'if ', 'for ', 'while ', 'try:', 'with ', :); \circled{2} Add the following code to the basic unit; \circled{3} Continue adding code until the next matched string is found; \circled{4} Extract a new basic unit if its size exceeds 4000 characters.
Each basic unit is self-contained and encapsulates specific behaviors or functionalities of the package.
This division ensures that: the length of each basic unit is manageable for the LLM; the code complexity is reduced; and each unit remains meaningful for rule generation. 
Breaking down the code into basic units enables the LLM to efficiently analyze and generate rules without being constrained by input length or complexity.

The LLM audits the code snippet to determine whether the package exhibits potential malicious places, as outlined in Table~\ref{tab:unit}. 
\circled{1} Indicators of Compromise (IoC): Information that indicates a high probability of unauthorized access to the system, such as DNS requests or IP addresses.
\circled{2} File Operation: Detection of suspicious file read and write operations within the code.  
\circled{3} Network Activity: Identification of API calls or requests made to malicious servers or IP addresses.  
\circled{4} Encryption Function: Detection of functions used for evasion or obfuscation techniques within the code.  
\circled{5} Privilege Operation: Identification of operations related to privilege escalation.
\circled{6} Anti-debug/Anti-analysis Operation: Detection of functions designed to prevent sandbox environments or debuggers.  

\textbf{Multiple Similar Units}.
After partitioning, the LLM audits multiple similar basic units irather than a single one.
This approach ensures that the generated rules are scalable and general, avoiding reliance on specific implementation details such as hardcoded strings or individual files.  
Several similar units are chosen from the same group.
While these units may exhibit slight differences in their code blocks, the LLM identifies and extracts common behaviors and features. 
This strategy enhances the rule's ability to generalize across various malicious patterns, improving its effectiveness and adaptability.

\begin{table}[!t]
    \centering
    \caption{The prompt in the LLM: instructions are used in the system role; \textcolor{blue}{Blue} indicates user input; \textcolor{orange}{Orange} indicates the (YARA$|$SemGrep) rule example; \{...\} indicates omitted content due to the page limitation.}
    \begin{tabular}{p{8.3cm}}
    \toprule
    \textbf{Prompt on rule generation from the basic unit.} \\
    \midrule

    \textbf{System role is as follows:}
    
    \textbf{Task.} 
As a senior malware code analyst, please analyze the following code samples from the same malware cluster and design effective \textcolor{orange}{\{YARA$|$SemGrep\}} rules. These samples are variants from the same malware family.

Sample 1: \textcolor{blue}{\{user input\}}

Sample 2: \textcolor{blue}{\{user input\}}
    
 \textbf{Thought Process:}
 
1. Initial Analysis: \{ ... ... \} \

2. In-depth Analysis: \{ ... ... \} \textit{refer to Table~\ref{tab:unit}}

3. External Knowledge Analysis: \{ ... ... \}

4. Understanding and Validation: \{ ... ... \}

\textbf{Output.} 

1. Analysis Result \{*.txt format\}

2. Write \textcolor{orange}{\{YARA$|$SemGrep\}} rules based on the analysis result. \\


    \midrule

    \textbf{User's information is as follows:}


    \textcolor{blue}{Input: \{Basic Unit One\}}
    
    \textcolor{blue}{Input: \{Basic Unit Two\}} 

   \textcolor{orange}{Few Shot: \{rule file\} }  \\

    
    \bottomrule
    \end{tabular}
    
    \label{tab:prompt}
\end{table}

\textbf{Prompt} plays an important role in guiding the LLM to generate the expected results through a series of instructions.  
Table~\ref{tab:prompt} lists the prompt used for the basic unit rule creation. 
The prompt follows the Chain-of-Thought (CoT) methodology, which divides the task into a series of linear steps.  
Specifically, the task is divided into 4 steps: initial analysis, in-depth analysis, external knowledge analysis, and validation.
(1) Initial analysis: Perform a code audit on the basic unit and provide a summary of the code.  
(2) In-depth analysis: Extract features or strings from the code based on the criteria listed in Table~\ref{tab:unit}.
(3) External knowledge analysis: Determine whether the input matches known malicious behavior patterns, such as worm propagation, ransomware encryption, or remote command execution. If a match is identified, existing patterns are leveraged to construct a rule. 
(4) Validation: Ensure reasoning consistency and confirm that the rule covers the potential behaviors exhibited by the code.

\textbf{Output}.
This component in {\tech} produces two outputs.
The first output is a detailed analysis result, saved in the `*.txt` format. 
This file provides a summary of the insights into the thought process.  
The second output is a rule in either the YARA or Semgrep format.
To guide the LLM, we leverage the few-shot learning technique by providing correct rule formats as references during rule generation. 
It is important to note that despite these measures, LLMs still introduce errors or hallucinations in the generated outputs.

\subsection{Refining Rules}

Two outputs from the previous stage (coarse-grained rules and analysis results) serve as the inputs for this task.  
The prompt (Table~\ref{tab:prompt:2}) provides the guidelines for this process.  
Specifically, the task is divided into 2 steps: rule analysis and rule optimization.

\textbf{Self-reflection}. 
We employ the self-reflection technique to guide the LLM in auditing coarse-grained rules. This approach leverages the LLM's capability to evaluate and critique its own prior outputs, enhancing overall performance~\cite{shinn2023reflexion, liu2023languages}.  
In the context of {\tech}, the self-reflection component ensures that the rules align with the analysis results. If discrepancies or inconsistencies are identified, the LLM revises the rules accordingly to maintain alignment and accuracy.

\begin{table}[!t]
    \centering
    \caption{The prompt in the LLM: instructions are used in the system role.}
    \begin{tabular}{p{8.3cm}}
    \toprule
    \textbf{Prompt on refining rules.} \\
    \midrule
    
\textbf{System role is as follows:}
    
    \textbf{Task.} 
You are a \{YARA$|$SemGrep\} rule expert. Your task is to analyze and optimize the input rules. Please follow these steps to ensure the rules are complete and efficient:

Analysis result: \textcolor{blue}{\{user input\}}

Rule: \textcolor{blue}{\{user input\}}

 \textbf{Thought Process:}
 
1. Self-reflection: \{ ... ... \}  

2. Optimize Rules: \{ ... ... \} 


     \textbf{Output:}

     \textcolor{orange}{\{YARA$|$SemGrep\}} rules 
    \\
    \midrule

    \textbf{User's information is as follows:}

    \textcolor{blue}{Input: \{Analysis Result\}} refer to Section~\ref{sec:task1}
    
    \textcolor{blue}{Input: \{YARA$|$SemGrep rule\}} refer to Section~\ref{sec:task1} \\

    \bottomrule
    \end{tabular}
    \label{tab:prompt:2}
\end{table}

\textbf{Rule Optimization}. 
During this step, coarse-grained rules are optimized and merged into a single, scalable rule. This process ensures that the resulting rule is both effective and general enough to detect a broader range of malicious behaviors while maintaining efficiency.  
Several guidelines are followed in this subtask:  
\begin{enumerate}  
    \item The \texttt{string} section should encapsulate malicious behaviors, such as API calls, file operations, and network activities.  
    If this is not the case, the \texttt{string} section should be revised based on the analysis results.  

    \item Standard naming conventions are applied to the \texttt{string} section to enhance consistency and readability.  
    When the \texttt{string} sections of multiple rules show similarities, regular expressions are used to manage potential variants.  

    \item Logical combinations (\texttt{all of them}, \texttt{any of them}, or regular expressions) are employed to merge rules.  
    If two rules overlap in scope, the rule with smaller coverage is removed.  

    \item The combined rule adheres to the required structure and format.  
    A valid rule begins with the keyword \texttt{rule} followed by a unique identifier.  
    Each YARA rule contains three sections: \texttt{meta}, \texttt{strings}, and \texttt{condition}.  

    \item The LLM minimizes resource-intensive operations (e.g., regular expressions) to optimize rule execution, ensuring efficiency without unnecessary overhead.  
\end{enumerate}

\textbf{Output}.
This component outputs a calibrated rule in YARA or Semgrep format.



\subsection{Aligning Rules}



\begin{figure}[!t]  
    \centering
    \includegraphics[width=1.0\linewidth]{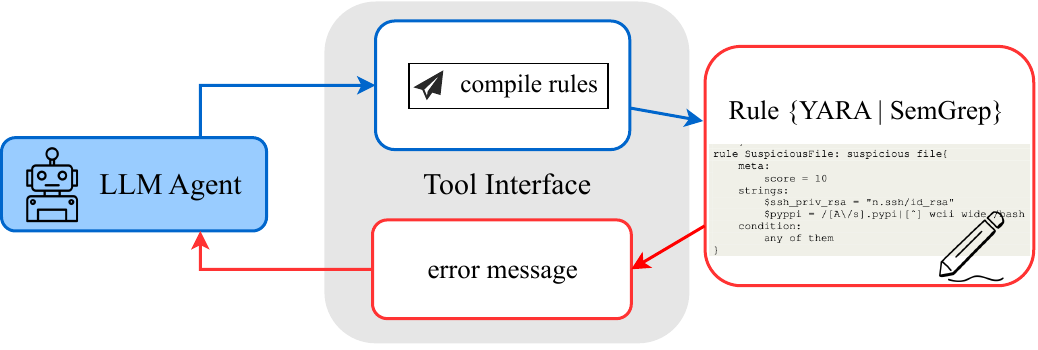}
    \caption{Error correction module: an agent-based LLM fixes errors based on a tool interface.}
    \label{fig:agent}
\end{figure}

Due to the inherent limitations of LLM, errors or hallucinations in generated rules are inevitable.
To address this, we propose an LLM-based agent designed to correct rule errors, as illustrated in Figure~\ref{fig:agent}.
Specifically, the agent utilizes the tool to compile the rules.
A blue color indicates that the generated rule has passed the verification process, while red indicates that it has failed. If a rule fails, the agent refines it further by analyzing the error messages.

\textbf{Tool}.
Although LLMs are capable of reading and generating text or images, they cannot interact with external environments or perform tasks such as searching or code execution. 
To overcome this limitation, we equip the agent with tools that enable communication between the LLM and the rule compiler.
Specifically, we propose two manually created tools: one for compiling YARA rules and another for compiling Semgrep rules. 
The compiler does not raise an exception when a rule is correct; however, it will raise exceptions for issues such as syntax or compilation errors. These tools are implemented as a Python file, which can be stored and updated as needed.

\textbf{Memory} serves as the core of the agent, connecting tools, LLMs, and task execution. It is used to store short-term information during processing. Specifically, error messages generated by the rule compiler are treated as observations and stored in memory. The agent retrieves these error messages and feeds them into the task execution component. Each time a rule compilation fails, new error messages are generated. If a rule fails to compile multiple times, the memory history can grow excessively. To address this, the memory module retains only the two most recent compilation error messages.

\textbf{Prompt}. 
We design a prompt to guide the LLM-based agent in fixing rules, as detailed in Table~\ref{tab:prompt:3}. 
 The agent’s tool generates error messages, which {\tech} uses to address and correct rule issues.
 {\tech} adheres strictly to the requirements and error messages to refine the rules, following these steps:
(1) Ensure the rule contains all necessary components: `meta', `strings', and `condition'.
(2) Check for syntax issues such as unmatched brackets, unclosed quotes, and other errors.
(3) Verify that all strings referenced in the condition section are properly defined in the `strings' section.
(4) Validate the correctness, efficiency, and expected matching behavior of all regex patterns.
(5) Confirm that the meta fields are well-formatted and meaningful.
(6) Ensure the rule is UTF-8 encoded without a BOM and that line endings align with the target environment.
If a rule successfully compiles, {\tech} outputs the corrected rule. Otherwise, it attempts to fix the rule up to five times.

\begin{table}[!t]
    \centering
    \caption{The prompt in the LLM: 6 instructions are used in the system role; \textcolor{red}{Red} indicates error messages from the agent's observations.}
    \begin{tabular}{p{8.3cm}}
    \toprule
    \textbf{Prompt on the rule fix.} \\
    \midrule
        
\textbf{Task.}
You are a \{YARA$|$SemGrep\} rule expert. Your task is to fix and optimize the input YARA rules. Please follow these steps to ensure the rules are complete, syntactically correct, and efficient:

Error message: \textcolor{red}{\{error\_info\}}

Analysis result: \textcolor{blue}{\{user input\}}

Rule: \textcolor{blue}{\{user input\}}

\textbf{Instruction.}

1. Missing or Incomplete Parts:   \{ ... ... \} \

2. Syntax Errors:  \{ ... ... \} 

3. Undefined Strings in Conditions:  \{ ... ... \} \

4. Regular Expression Issues:  \{ ... ... \} \

5. Invalid `meta' Field Values:  \{ ... ... \} \
   
   6. File Encoding Issues:  \{ ... ... \} \
   \\
    \midrule

   \textbf{User's information is as follows:}

    \textcolor{blue}{Input: \{Analysis Result\}}
    
    \textcolor{blue}{Input: \{YARA$|$SemGrep rule\}} \\
     \midrule
    \textbf{Agent's observation is as follows:}

    \textcolor{red}{Error: \{error result\}} \\
    \bottomrule
    \end{tabular}
    \label{tab:prompt:3}
\end{table}


\section{Evaluation}
\label{sec:exp}

In this section, we conducted a series of experiments to validate the effectiveness of {\tech} in generating rules for detecting malicious packages. 
Then, we provided a systematic analysis of those generated rules.



\begin{table}[!t] 
    \caption{The details of the dataset for OSS malicious packages. }
    \label{tab:data}
    \centering
    \begin{tabular}{  c c c c   }
    \toprule
    Category                  &  Pkg. Num.   &  Deduplicated  Num.  & Avg. LoC     \\
    \toprule
    Malware            &  3,200  &   1,633   &   424   \\
    Legitimate        &   500    &   500     &   3,052 \\
    \toprule    
    \end{tabular}
\end{table}

\subsection{Experimental Setting}

\textbf{Implementation}.
We have implemented a prototype system of {\tech} via several open-source libraries. 
{\tech}'s input is a malicious package, and the output is its corresponding YARA \& Semgrep rules. 
{\tech} uses regular expression matching to extract the package metadata, where we use the re library~\cite{regex} to implement the regex. 
{\tech} use the tokenize~\cite{nltk} library to convert source code to tokens, and the CodeBERT embedding model to generate code's vectors. 
We leverage the Scikit-learn~\cite{scikit} library to implement the clustering algorithm to divide code fragments.
{\tech} uses the LangChain~\cite{langchain} module to automatically generate rules based on the extracted metadata and code fragments.

\textbf{Dataset}.  
Table~\ref{tab:data} lists the details of the dataset, including 3,200 malware packages and 500 legitimate packages. 
The malware comes from the GuardDog~\cite{guarddog} that provides PyPI malicious packages to the public via the GitHub repository.
The legitimate packages come from the most popular PyPI packages (by download count) over one year~\cite{pkg-popular}.
We find there are many duplicate packages in the malware dataset, where their signatures are the same. 
After deduplicated, the number of malware is 1,633. 
The average number of malware code (LOC, Line of code) is 424 lines.
For the legitimate packages, the average number of code segments is 3,052 LoC.
It is obvious that the number of code segments in the malware packages is much smaller than that in the legitimate packages. 
The reason is straightforward: malware packages are designed for a specific purpose, such as data theft, backdoor access, or surveillance, and don't require extensive functionality.

\begin{table}[!t] 
    \caption{The details of baselines. }
    \label{tab:baselines}
    \centering
    \begin{tabular}{  l c     }
    \toprule
     Category &  \textbf{Method}     \\
    \toprule
    \makecell[l]{Existing Rules\\ from SOTA Tools} & \makecell[c]{Yara scanner~\cite{naik2020evaluating},  Semgrep scanner~\cite{semgrep}}  \\\hline
    \multirow{1}{*}{Score-based Approach} &   Prior works~\cite{raff2020automatic, brengel2021yarix}      \\ \hline
    \multirow{1}{*}{ Diverse LLMs} &  \makecell[c]{GPT-3.5 turbo, GPT-4o~\cite{openai} \\ Claude-3.5-Sonnet~\cite{claude}, Llama-3.1 70B~\cite{llama} }      \\
    \toprule
    \end{tabular}
\end{table}

\textbf{Baselines}. 
We compare {\tech} with several baselines: SOTA tools, automatic rule generation, and diverse LLMs. Table~\ref{tab:baselines} lists the details of the baselines.

\textit{(1) Existing Rules from SOTA Tools.}  
The first category involves existing rules from YARA~\cite{naik2020evaluating} and Semgrep scanners~\cite{semgrep}.
YARA scanner uses 4,574 YARA format rules to find risks and threats, covering vulnerabilities, malware, shells, mobile applications, emails, etc.  
Semgrep scanner uses 2,841 Semgrep format rules to find risks and threats, covering third-party software, cloud services, network communications, systems, network applications, etc.
Those rules were written by developers, security professionals, or researchers.
We use those rules to compare the effectiveness of rules generated by {\tech}.


\textit{(2) Score-based Approach}. 
So far, there is no approach for generating rules to target OSS malware.
Instead, several score-based approaches~\cite{raff2020automatic, brengel2021yarix} and tools~\cite{vovk} can generate signatures from binary files. 
Hence, we revise the score-based approach to adapt to OSS malware.
(1) First, we use 3 types of scores to measure the importance of strings: isolation forest, information entropy, and TF-IDF (Term Frequency-Inverse Document Frequency).  
Each score is assigned a specific weight: isolation forest is given a weight of 1.2, TF-IDF has a weight of 1.0, and information entropy is weighted at 0.8.
(2) Next, we apply a clustering algorithm (Section~\ref{sec:sub:code}) to partition both malware and legitimate packages (Table~\ref{tab:data}) into different code groups.
(3) In each iteration, we pick up two groups (one from malware and one from legitimate) and use 3 scores to calculate the importance of strings between the two groups.
(5) Strings with high scores (above a 0.9 threshold) are picked up into the `string' part of the YARA format rule.
The remaining parts of the rules are generated through a rule template.


\textit{(3) Diverse LLMs}. 
We use several large language models (LLMs) to generate rules, including GPT-3.5 turbo, GPT-4o~\cite{openai},  Claude-3.5-Sonnet~\cite{claude}, Llama-3.1:70B~\cite{llama}.
The first three LLMs belong to online services, and Llama-3.1 belongs to the local LLM.
We leverage APIs of online LLMs to generate rules. 
For Llama-3.1-70B, we deploy it on the local server. 
We inspect the diverse efforts of different LLMs on rule generation.

\subsection{Performance}

\begin{table}[!t] 
    \caption{ Performance of {\tech} compared to baselines. }
    \label{tab:rulellm:classification}
    \centering
    \begin{tabular}{  l c c c c    }
    \toprule
    \textbf{Rule Type}          & \textbf{Accuracy} & \textbf{Precision} & \textbf{Recall} & \textbf{F1} \\
    \toprule
    \makecell[l]{\colorbox{gray}{\tech} }  & \colorbox{gray}{81.4\%}   & \colorbox{gray}{85.2\%}   & \colorbox{gray}{91.8\%}  & \colorbox{gray}{88.4\%}  \\
    Yara scanner       & 41.6\%   & 35.0\%    & 23.4\%  & 28.0\%  \\
    Semgrep scanner    & 56.2\%   & 70.9\%    & 32.0\%   & 44.0\%   \\
    Score-based   & 84.5\%   & 47.8\%   & 66.6\%  & 55.7\%  \\
    \toprule
    \end{tabular}
\end{table}

\textbf{Effectiveness of Rules.}
First, we evaluate the effectiveness of rules generated by {\tech} for detecting malicious and legitimate packages.
Table~\ref{tab:rulellm:classification} lists the comparative performance of rules generated by {\tech} against various baselines, including Score-based, YARA scanner, and Semgrep scanner. 
We use 4 metrics to represent the performance, including accuracy, precision, recall, and F1 score.
{\tech} achieved promising performance, with an accuracy of 81.4\%, precision of 85.2\%, recall of 91.8\%, and F1 score of 88.4\%, outperforming most other methods.  
Both scanners show significantly lower performance compared to {\tech}, with the Yara scanner performing the worst, achieving only 41.6\% accuracy and an F1 score of 28.0\%.
The Score-based method performs well in accuracy (84.5\%) but falls short on other metrics, indicating possible overfitting or reliance on specific criteria that may not generalize well.
{\tech} demonstrates superior performance compared to the baselines, as evidenced by its high recall, precision, and F1 score. 
It effectively balances the identification of true positives and minimization of errors.

\begin{figure}
    \centering
    \includegraphics[width=2.4in]{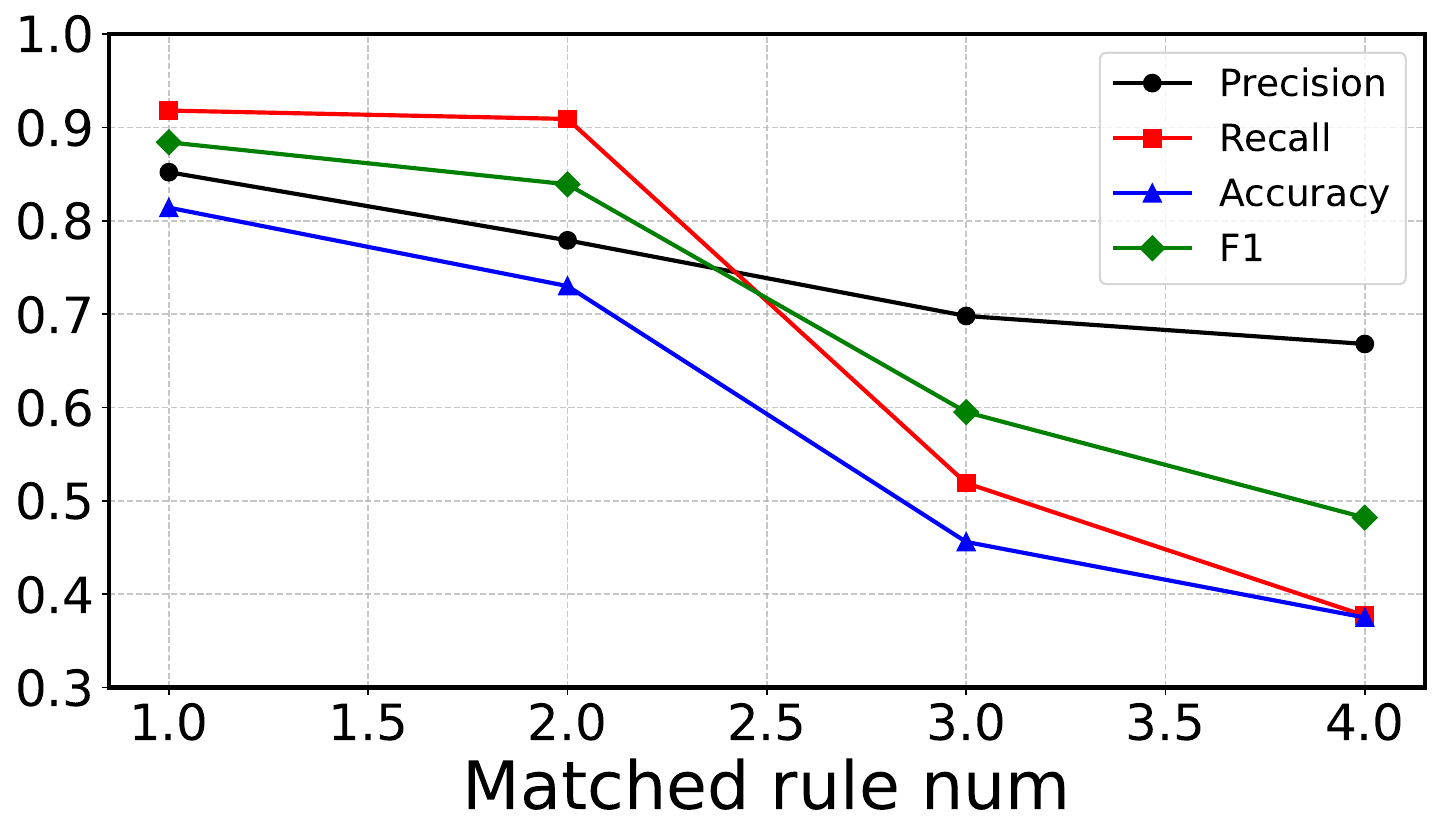}
        \caption{YARA rule: the malware detection's performance along with the matched number. }
        \label{fig:perf:yara:num}
\end{figure}

\begin{figure}
    \centering
    \includegraphics[width=2.4in]{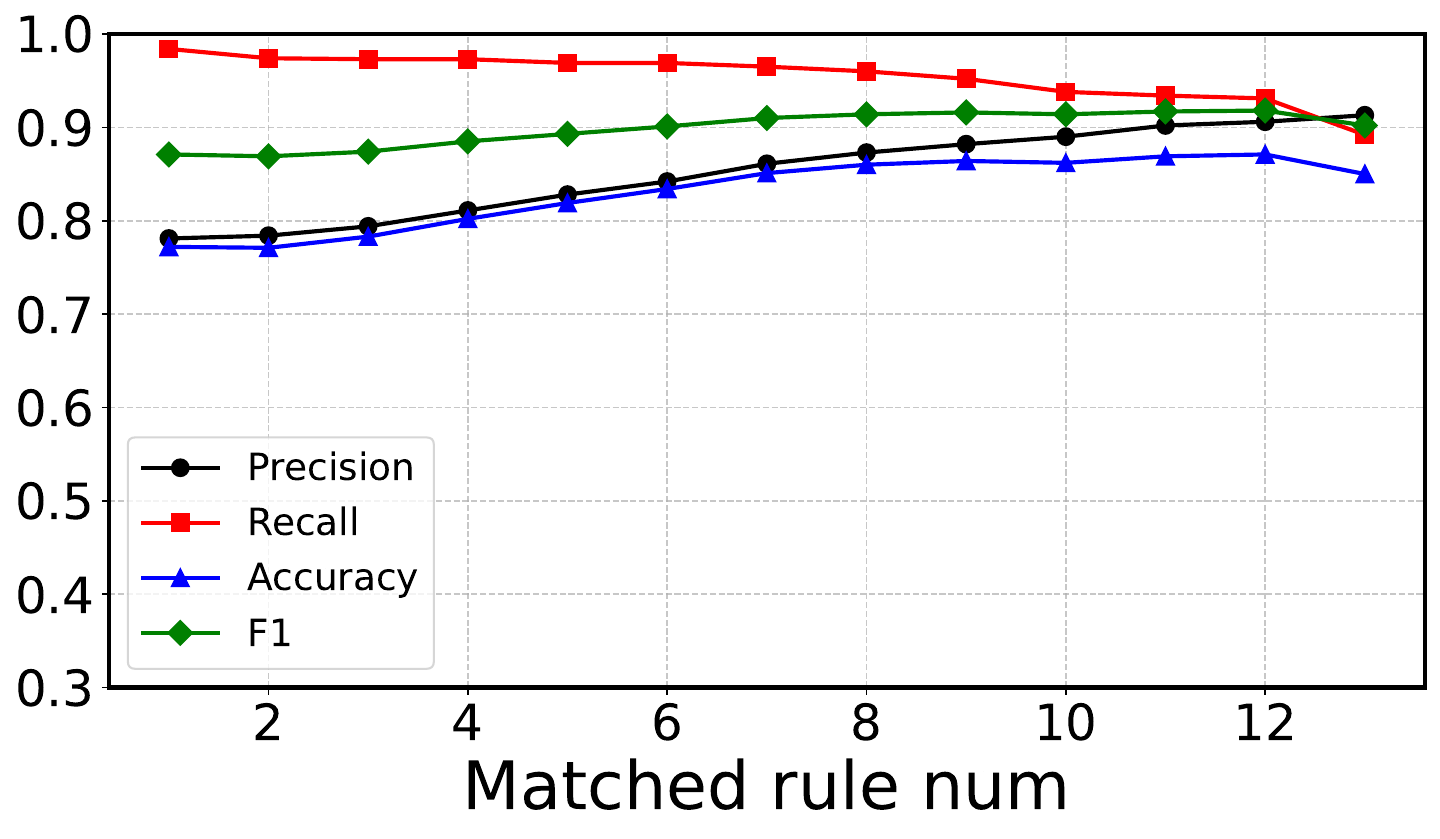}
        \caption{Semgrep rule: the malware detection's performance along with the matched number.  }
        \label{fig:perf:sem:num}
\end{figure}

We further inspect the malware detection performance along with the number of matched rules.  
Figure~\ref{fig:perf:yara:num} depicts the performance distribution (accuracy, precision, recall, and F1) of YARA rules, with the X-axis representing the number of matched rules. 
It is evident that when the matched rule number is equal to 1, the malware detection achieves the best performance.
In addition, performance decreases continuously as the number of matched rules increases.  
This is because YARA-generated rules are highly specific and do not share similar patterns.  
Figure~\ref{fig:perf:sem:num} shows the performance of Semgrep rules relative to the number of matched rules.
We observe that the performance of Semgrep-generated rules changes only slightly with the number of matched rules. 
When the number of matched Semgrep rules is equal to 9, malware detection reaches its peak performance. 
This can be attributed to the fact that Semgrep rules focus on code structures and static analysis, while YARA rules are centered on signatures and specific patterns.
Hence, the Semgrep rules have broader patterns than the YARA rules. 
Overall, the number of matched rules (Figures~\ref{fig:perf:yara:num} and \ref{fig:perf:sem:num}) in malware detection demonstrates that {\tech} can effectively generate both YARA and Semgrep rules.

Table~\ref{tab:llm:performance} compares the performance of rules generated by various LLMs, including GPT-3.5-turbo, GPT-4o, Claude-3.5-Sonnet, Llama3.1:70b.
GPT-4o achieves the highest performance across all metrics, with an accuracy of 81.4\%, precision of 85.2\%, recall of 91.8\%, and an F1 score of 88.4\%. 
Both Llama3.1:70b and GPT-3.5-turbo show moderate performance, with an accuracy of 72.6\% and 74.5\%.
Notably, Claude-3.5-Sonnet's precision is lower at 75.0\%, though its recall is relatively higher at 95.9\%.
These results suggest that GPT-4o excels at generating rules that are both precise and effective, capturing a high number of true positives and maintaining low false positives.

\begin{table}[!t] 
    \caption{Performance of Rules Generated by different LLMs. }
    \label{tab:llm:performance}
    \centering
    \begin{tabular}{  l c c c c    }
    \toprule
    \textbf{Rule Type} & \textbf{Accuracy} & \textbf{Precision} & \textbf{Recall} & \textbf{F1 Score} \\
    \toprule
    GPT-3.5 turbo    & 72.6\%   & 78.4\%   & 68.0\%  & 72.8\%  \\
    GPT-4o       & \colorbox{gray}{81.4\%}   & \colorbox{gray}{85.2\%}   & \colorbox{gray}{91.8\%}  & \colorbox{gray}{88.4\%}  \\
    Claude-3.5-Sonnet     & 75.0\%   & 77.3\%   & 95.9\%  & 85.6\%  \\
    Llama-3.1:70B  & 78.2\%   & 68.0\%   & 72.6\%  & 77.4\%  \\
    \toprule
    \end{tabular}
\end{table}

\textbf{Malware Variant Detection}.
We further inspect whether rules generated by {\tech} can detect variants of OSS malware. 
We use the clustering algorithm (Section~\ref{sec:sub:code}) to divide malware packages into different groups.
In each group, we use two malware packages to generate YARA rules, and the rest packages are unknown variants.
We use those generated rules to detect unknown variants in the same group.
The overall detection rate is 90.32\%, and the average detection rate is 96.62\%.
The results demonstrate that rules generated by {\tech} can detect potential variants.

\begin{table}[!t] 
    \caption{Ablation Experiment: impact of each component's effectiveness in {\tech}. }
    \label{tab:eval:com}
    \centering
    \begin{tabular}{  l c c    }
    \toprule
    \textbf{Approach} & \textbf{Precision} & \textbf{Recall} \\
    \toprule
    LLMs alone & 62.9\%   & 56.8\%  \\
    \midrule
    LLM + Rule Alignment  &  79.2\% &  84.3\%   \\  \hline
    \makecell[l]{LLM + Basic-unit Rule + \\ Rule Alignment }  & 81.90\% & 90.0\%    \\\hline
    \makecell[l]{ LLM + Basic-unit Rule + \\ Combination + Rule Alignment}          & \colorbox{gray}{85.2\%}   & \colorbox{gray}{91.8\%}    \\ 
    \toprule
    \end{tabular}
\end{table}

\textbf{Ablation Experiment}.
We validate the impact of each component in {\tech} (Figure~\ref{fig:arch:llm}): basic-unit rule creation, rule combination, and rule alignment.
Specifically, we use 4 approaches: (1) LLMs alone; (2) LLM + rule alignment; (3) LLM + basic-unit rule + rule alignment; and (4) LLM + basic-unit rule + combination + rule alignment.
To ensure a fair comparison, all prompts and their requirements (Table~\ref{tab:prompt}-\ref{tab:prompt:3}) are consistently used in LLM.
Table~\ref{tab:eval:com} lists each component's effectiveness in {\tech}. 
Directly using LLM to generate rules leads to a significantly low recall (56.8\%) and relatively moderate precision (62.9\%).
Without the {\tech}'s help,  LLMs struggle with the rule generation task, missing a substantial portion of the rules. 
LLM + Rule Alignment can significantly improve both the precision and recall of generated rules. 
The reason is that the rule alignment can fix errors in rule formats, and this approach can find more useful rules.  
LLM + Basic-unit Rule + Rule Alignment yield a substantial increase in recall, jumping to 90.0\%.
The basic unit rules help this approach capture a broader range of true positives. 
{\tech} (a combination of various components) shows the highest performance across both precision and recall.
Thus, {\tech}’s effectiveness is not merely a result of using an advanced LLM; it stems from a combination of specialized techniques, optimizations, and a tailored agent-based architecture that improves rule quality.

\subsection{In-depth Analysis}

\begin{table}[!t]
    \caption{The rule number between {\tech} and SOTA tools.}
    \label{tab:rule:number}
    \centering
    \begin{tabular}{c c c c}
    \toprule
    \multirow{2}{*}{\textbf{Category}} & \multicolumn{2}{c}{SOTA Tool} & \multirow{2}{*}{{\tech}} \\
    \cmidrule(lr){2-3}
    & All Rules& OSS Malware & \\
    \midrule
    Yara Rule Format & 4,574 & 46 & \colorbox{gray}{452} \\
    Semgrep Rule Format & 2,841 & 334 & \colorbox{gray}{311} \\
    \bottomrule
    \end{tabular}
\end{table}

We provide an in-depth analysis of the rule quality, including the rule number, the precision per rule, and the coverage per rule.


\textbf{Rule number}.
Table~\ref{tab:rule:number} shows the rule number comparison between {\tech} and SOTA tools. 
{\tech} can generate two types of rules: Yara format and Semgrep format.
We can see that {\tech} has 452 rules in the Yara rule format and 311 rules in the Semgrep rule format.
In comparison, SOTA tools have 4,574 total rules and 46 OSS malware rules in the Yara format (2,841 total rules and 334 malware rules in the Semgrep format). 
Rules from tools (Semgrep and Yara scanners) were written by security experts, requiring domain-specific knowledge and manual efforts. 
In addition, most of the YARA and Semgrep rules are not designed for OSS malware. 
Yara scanners are designed for signatures and are difficult to deal with malware deformation, and Semgrep scanners usually support taint analysis and string matching.
Although the rule amount from {\tech} is smaller than that of SOTA tools, {\tech} shows the best performance (detailed in Table~\ref{tab:baselines}).
Rules generated by {\tech} have broader detection coverage to recognize a variety of patterns, behaviors, or anomalies in malicious packages.

\begin{figure}[!t]
    \centering
    \includegraphics[width=2.4in]{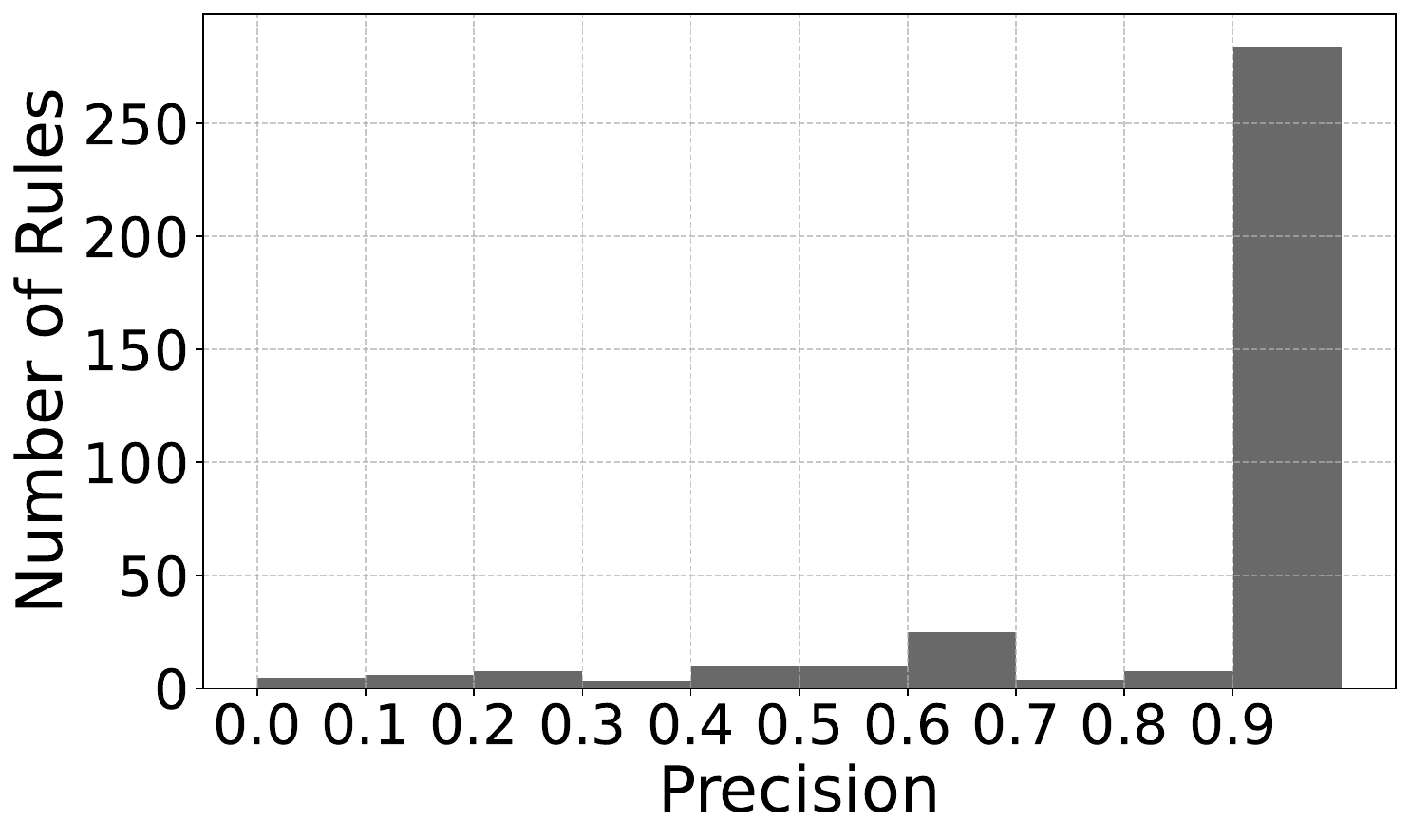}
        \caption{YARA rule's precision: the precision distribution for all rules generated by {\tech}. }
        \label{fig:bar:perf:yara}
\end{figure}

\textbf{Precision per rule}. 
We inspect the precision performance of every rule generated by {\tech}. 
Figure~\ref{fig:bar:perf:yara} depicts the distribution of 452 YARA rules' performance, where the X-axis is precision, and the Y-axis is the rule number. 
It is observed that 278 YARA rules have a high precision, nearly 98.2\%, and the rest rules have various precisions. 
For each rule, the high precision indicates a high confidence in malware and threat detection.
If a package matches a rule with high precision, it is confident that the package is malicious.
Note that 65 YARA rules (452-387) do not match with any malicious packages.

\begin{figure}[!t]
    \centering
    \includegraphics[width=2.4in]{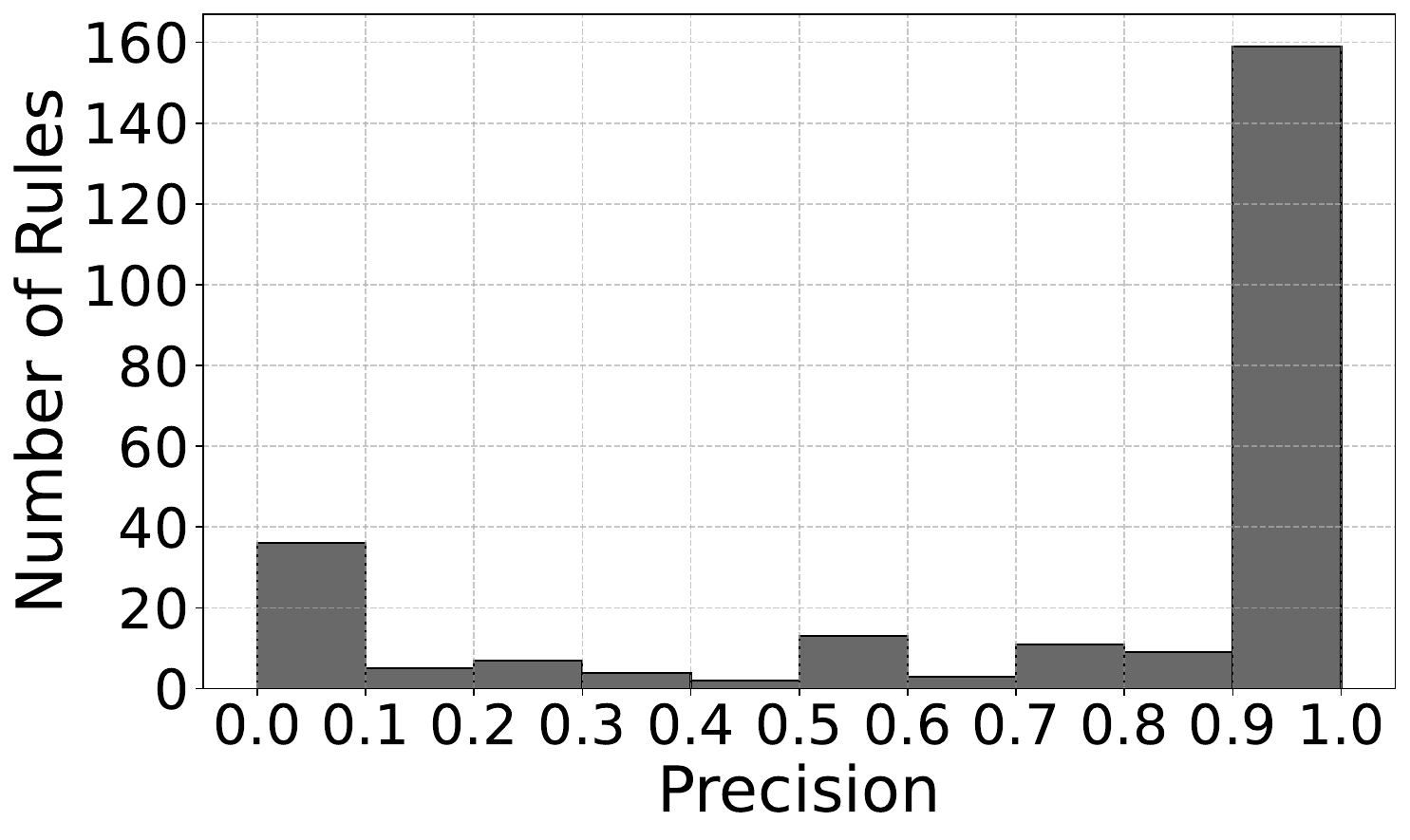}
        \caption{Semgrep rule's precision: the precision distribution for all rules generated by {\tech}. }
        \label{fig:bar:perf:sem}
\end{figure}

Figure~\ref{fig:bar:perf:sem} depicts the distribution of 311 Semgrep rules' performance.
Similarly, 158 Semgrep rules have a high precision, nearly 97.1\%.
We also find that nearly 40 Semgrep rules have close 0\% precision and 62 Semgrep rules (311-249) do not match with any malicious packages. 
In terms of manual inspection, those rules use a highly specific taint-code structure, leading to poor performance.

\begin{figure}[!t]
    \centering
    \includegraphics[width=2.4in]{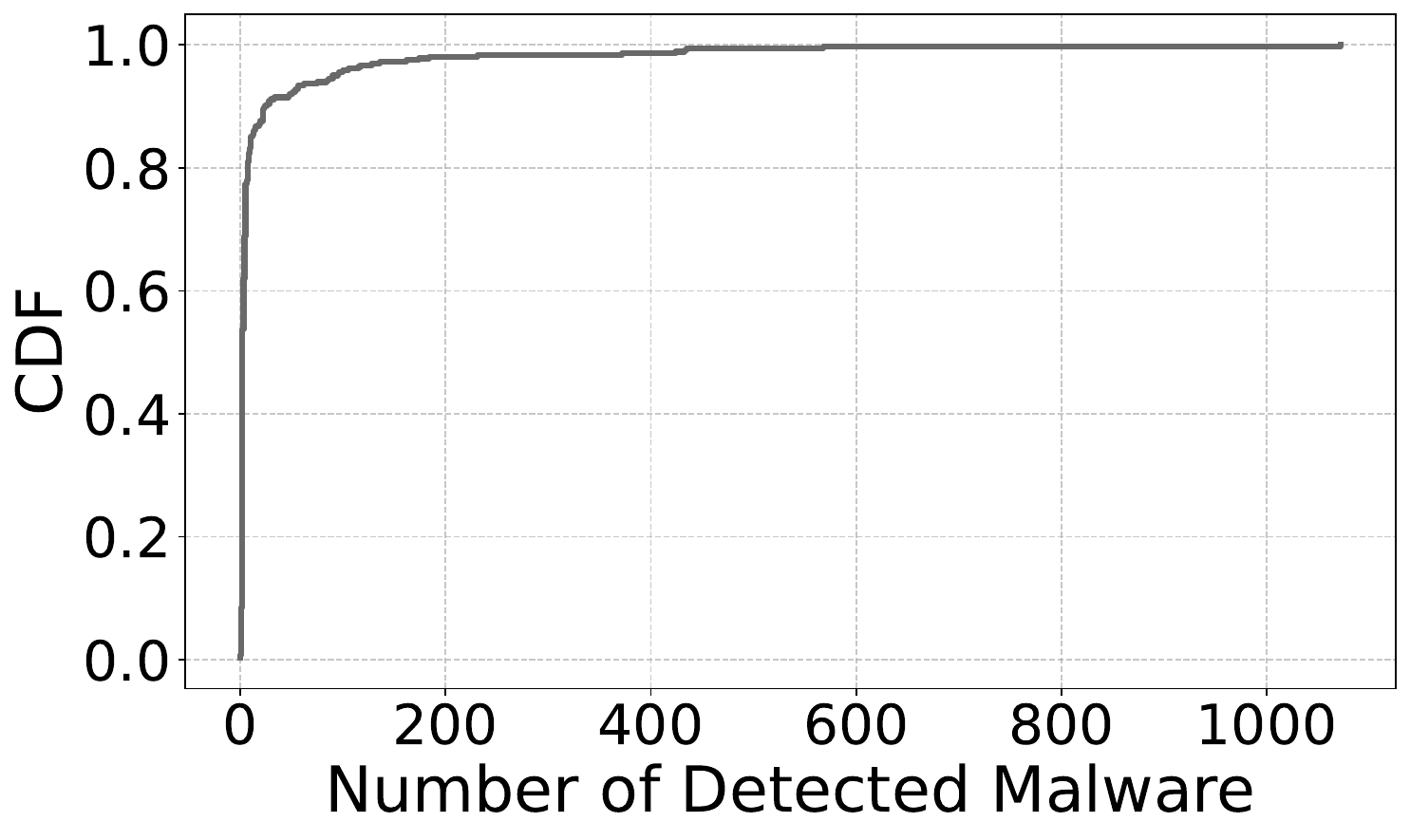}
        \caption{YARA rule's coverage: the CDF of the detected malware per rule generated by {\tech}.}
        \label{fig:cdf:malware:yara}
\end{figure}

\textbf{Coverage per rule}.
Next, we use the number of detected malware packages to reflect the coverage of rules generated by {\tech}.  
A high number of detected malware packages indicates that a rule has broad coverage and identifies common patterns, while a low number of detected malware packages suggests that a rule is more specific and suited to a narrow set of patterns. 
Figure~\ref{fig:cdf:malware:yara} shows the CDF of the detected malware number for YARA rules. 
It is evident that many YARA rules detect a small number of malware packages, whereas 80\% rules cover fewer than 10 packages. 
However, 10 YARA rules detect over 100 malware packages, indicating broader and more common patterns.  
For example, a rule related to the fake version can detect 568 malware packages and a rule related to the C2 server can detect 185 malware packages.
In short, most YARA rules use a specific string or regex, minimizing false positives by focusing on unique identifiers or signatures.

\begin{figure}
    \centering
    \includegraphics[width=2.4in]{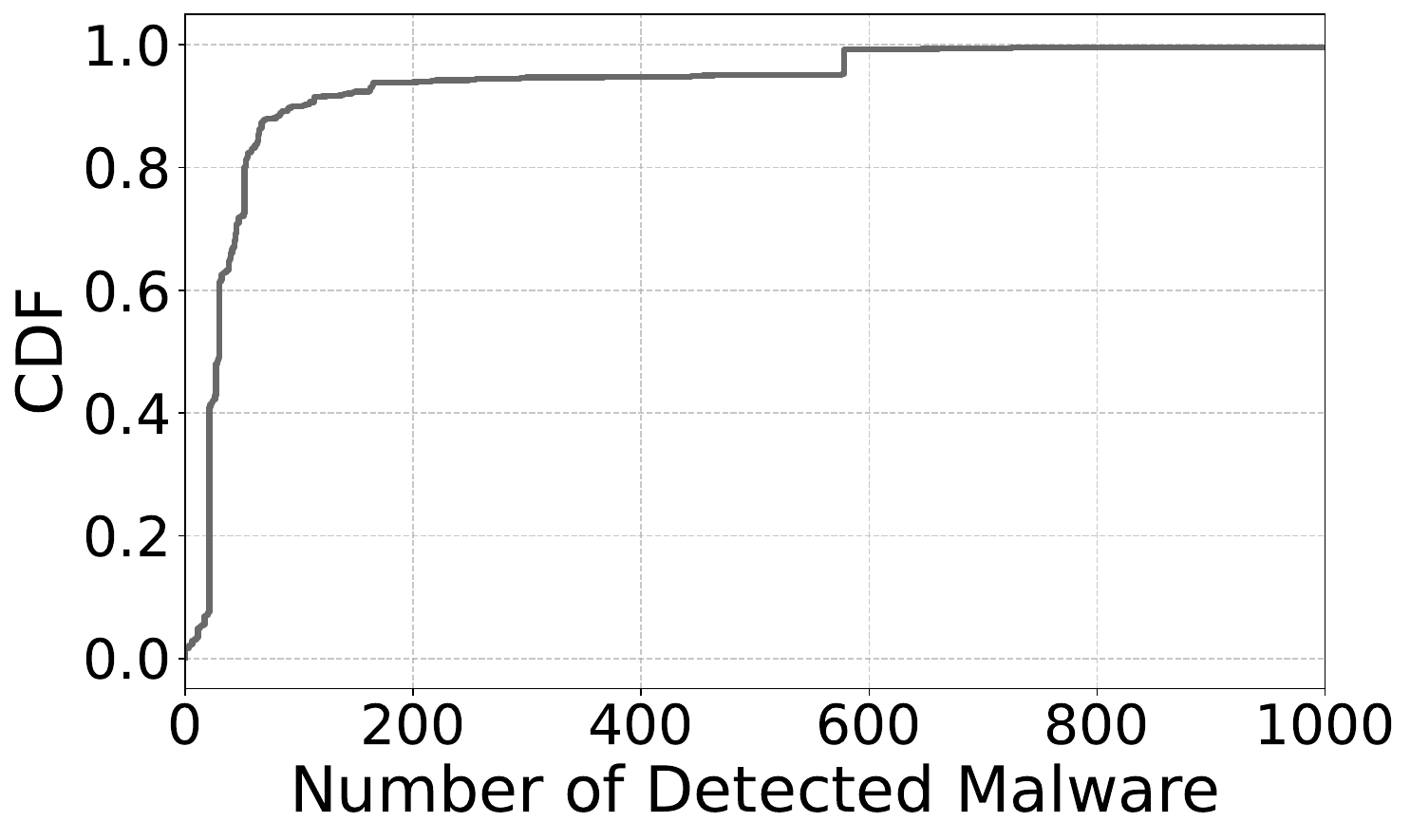}
        \caption{Semgrep rule's coverage: the CDF of the detected malware per rule generated by {\tech}.}
        \label{fig:cdf:malware:semgrep}
\end{figure}

Figure~\ref{fig:cdf:malware:semgrep} shows the CDF of detected malware number for Semgrep rules. 
It is observed that Semgrep rules have a broader range for detecting malware packages compared to YARA rules.  
Only 40\% of Semgrep rules cover fewer than 10 malware packages, whereas other Semgrep rules with broader patterns may result in a higher false positive rate.  
In contrast, broad Semgrep rules have a high recall (detailed in Figure~\ref{fig:perf:sem:num}).
Semgrep rules can be advantageous for identifying a broad spectrum of threats at the initial stage.

Overall, YARA rules differ from Semgrep rules in their matching patterns and targets.  
A broad rule is useful for a quick scan but risks false positives, whereas a specific rule excels in precision but may miss underlying threats.

\subsection{Rule Analysis}

\begin{table}[!t]
    \caption{YARA rules: detailed breakdown of rule categories and subcategories}
    \label{tab:rule:subcategory}
    \centering
    \begin{tabular}{ l l r }
    \toprule
    Category Name & Subcategory Name & Count \\
    \midrule
    \multirow{4}{*}{0. Metadata Related} & Package Metadata Manipulation & 92 \\
    & Version Number Deception & 17 \\
    & Fake Dependency Metadata & 18 \\
    & Author Information Spoofing & 29 \\
    \midrule
    \multirow{4}{*}{1. Malicious Behavior} & Privilege Escalation & 21 \\
    & Process Manipulation & 25 \\
    & System Configuration Changes & 70 \\
    & Persistence Mechanisms & 87 \\
    
    \midrule
    \multirow{4}{*}{2. Dependency Library} & System Library Abuse & 25 \\
    & Network Library Misuse & 43 \\
    & Crypto Library Exploitation & 7 \\
    & UI/Graphics Library Abuse & 8 \\
    \midrule
    \multirow{4}{*}{3. Setup Code} & Malicious Setup Scripts & 56 \\
    & Build Process Manipulation & 11 \\
    & Installation Hook Abuse & 39 \\
    & Configuration Tampering & 28 \\
    \midrule
    \multirow{4}{*}{4. Network Related} & C2 Communication & 66 \\
    & Data Exfiltration Channels & 51 \\
    & Malicious Downloads & 61 \\
    & DNS/Protocol Abuse & 15 \\
    \midrule
    \multirow{4}{*}{\makecell[l]{5. Obfuscation \&\\ Anti-Detection}} & Code Obfuscation & 72 \\
    & Anti-Analysis Techniques & 67 \\
    & Sandbox Evasion & 9 \\
    & String/Pattern Hiding & 35 \\
    \midrule
    \multirow{4}{*}{6. Data Exfiltration} & Credential Theft & 8 \\
    & Environment Data Stealing & 31 \\
    & Configuration File Extraction & 2 \\
    & Sensitive Data Harvesting & 53 \\
    \midrule
    \multirow{3}{*}{7. Code Execution} & Shell Command Execution & 54 \\
    & Script Injection & 29 \\
    & Process Creation & 1 \\
    \midrule
    \multirow{4}{*}{8. Application} & Messaging Platform Abuse & 35 \\
    & Social Media API Exploitation & 2 \\
    & Cloud Service Misuse & 18 \\
    & Development Tool Abuse & 5 \\
    \midrule
    \multirow{2}{*}{9. Malware Family} & Known Trojan Families & 12 \\
    & Backdoor Families & 2 \\
    \midrule
    \multirow{1}{*}{10. Other Rules} & Unknown or Undetermined & 13 \\
    \bottomrule
    \end{tabular}
\end{table}

We manually inspect the content of generated rules and categorize them into 11 categories and 38 subcategories.  
The categorization relies on the nature of the rules and professional experience. 
Table~\ref{tab:rule:subcategory} provides a detailed breakdown of various rule categories and their respective subcategories, along with the count of rules in each subcategory. 
It is evident that the category with the highest number of rules is ``Malicious Behavior'' with a total of 203 rules, followed closely by ``Network Related'' with 193 rules.
``Malicious Behavior'' encompasses actions that directly harm the system or escalate privileges.
``Network Related'' includes behaviors that involve network communication, such as command and control (C2) communication or data exfiltration.
The high counts in categories underscore the complexity and prevalence of these threats from OSS malware.
Conversely, low counts in certain categories (e.g., ``Configuration File Extraction'')  may indicate less common or less complex threats.
In short, our generated rules highlight the diverse range of rule categories and subcategories used to detect and mitigate various security threats.

\begin{figure}
    \centering
    \includegraphics[width=0.7\linewidth]{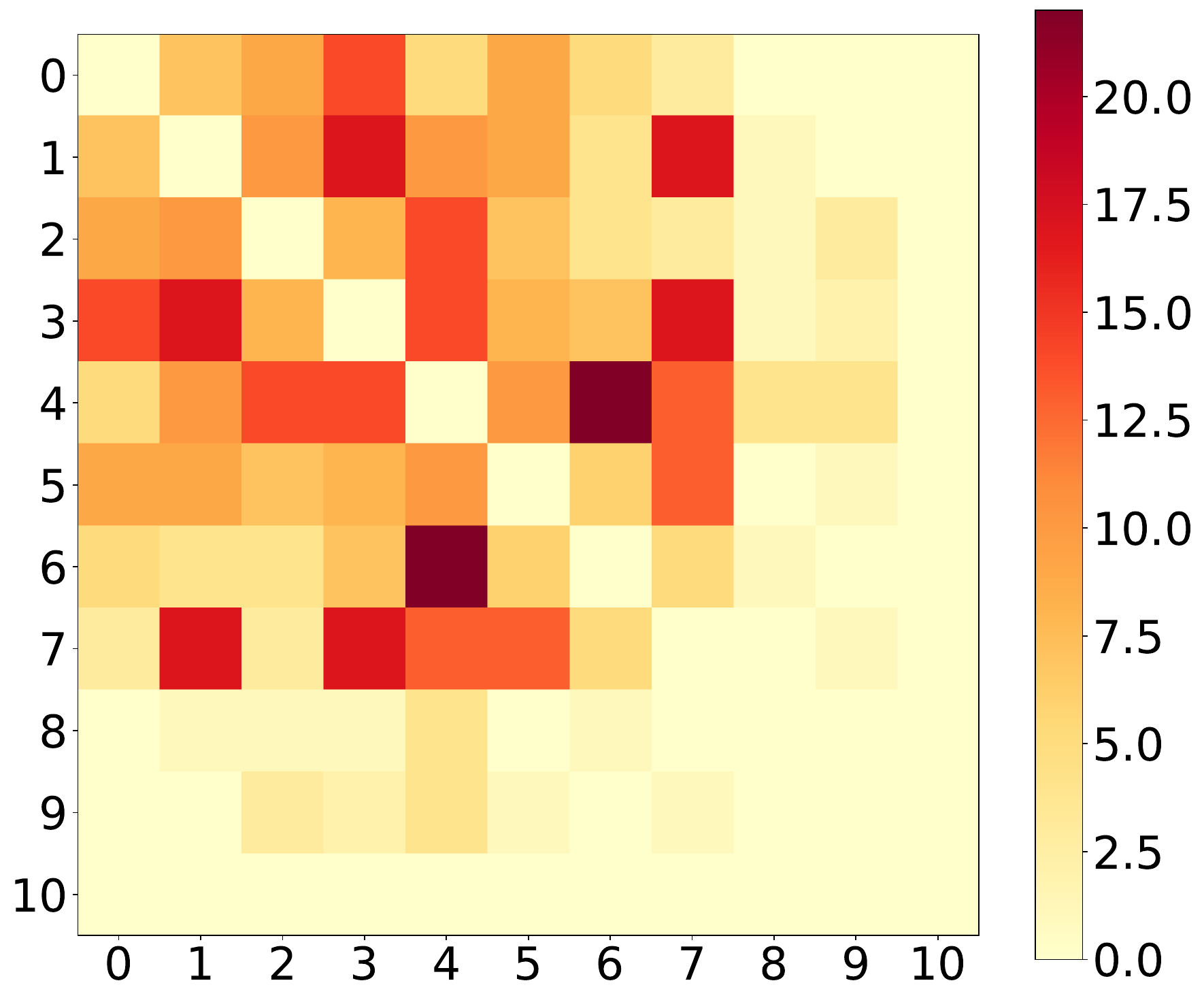}
    \caption{Heatmap: the overlapping degree between different categories; the number indicates the rule category in Table~\ref{tab:rule:subcategory}.}
    \label{fig:heat}
\end{figure}

\textit{(1) Non-exclusive category}. 
The rule category and subcategory are not mutually exclusive, and a rule can belong to multiple categories and subcategories. 
The total number of YARA rules in Table~\ref{tab:rule:subcategory} is 1,217, compared to 452 YARA rules in Table~\ref{tab:rule:number}.
Figure~\ref{fig:heat} depicts the overlapping rules among different categories.
For example, the `Malware\_Setuptools\_PostHook' rule covers two categories: ``Setup Code Rule' and ``Network Related Rule''. 
The reason is that {\tech} extracts rules from malicious packages, whereas a malware package may exhibit behaviors, e.g., a ransomware package might encrypt files while also exfiltrating data.

\textit{(2) Large Detection Range}. 
Several rule categories have a large detection range:  ``Obfuscation \& Anti-Detection'', ``Metadata-Related'', ``Code Execution'', ``Network Related'', and ``Data Exfiltration''. 
Rules in those categories have a large range, where more than 1,000 packages are detected. 
In the ``Code Execution'' category, a rule detected 23.41 malware packages on average; in the ``Obfuscation \& Anti-Detection'' category, a rule detected 19.50 malware packages on average.
It concluded that some techniques(e.g., common code snippets, network behaviors, or obfuscation patterns) are generic and widely in OSS malware.

\textit{(3) Narrow Detection Range}. 
Several rule categories have a narrow detection range: ``Malware Family'', and ``Application''. 
Rules in those categories have a small range, less than 5 packages per rule.
The reason is that those malware packages are less well-known.
Malware families are popular in conventional malware samples with different architectures and compilers, leading to binary polymorphism.
However, OSS malware is a software package to organize artifacts and components.  Malware families are not common in OSS malware. 
In brief, low-detection categories center around fewer but more specific matches of malware packages.




\section{Discussion \& Limitations}
\label{sec:discuss}


\textbf{Data Leakage}~\cite{xu2024benchmarkdatacontaminationlarge, shi2023detecting} leads to the concern of the {\tech}'s performance.
If the LLM uses the GuardDog dataset to pre-train the model with the same target, the validity of the performance in {\tech} may be overstated. 
There are 2 manners to mitigate the impact of data leakage of LLMs.
First, we can use malicious packages whose release time is newer than the cutoff date of model pre-training.
In our experiments, there are 78\% malicious packages whose release time is newer than Dec./2023, and GPT-4o's cutoff date is Oct./2023. 
There is a high probability that those malicious packages do not have data leakage issues. 
Second, even though LLM uses the GuardDog dataset to pre-train the model, the LLM still struggles with the rule generation task.
Our experimental results (Table~\ref{tab:eval:com}) show that directly using LLM to generate rules leads to a significantly low recall (56.8\%) and relatively moderate precision (62.9\%). 



\textbf{LLMs' Limitations}. 
{\tech}'s performance mainly relies on the reasoning ability of LLMs. However, as LLMs are trained on general datasets when handling specialized knowledge such as malicious packages, inaccuracies are inevitable. 
This can result in the generated rules being overly broad, leading to a high number of false positives, or only capable of identifying particular samples. 
In our {\tech}, the LLM can achieve 85.2\% precison. 
Another problem is the hallucinations caused by LLMs, where some rules may be fabricated content, or confuse truth and falsehood.
The reason for fabricated rules is that the LLM lacks a definitive ground truth for unknown malicious packages and unseen risks. 
Due to context length limitations, LLMs struggle to handle excessively long malicious code, such as obfuscated code or payload with base64 encoding.

\textbf{Retrieval Augmented Generation} (RAG) can provide external knowledge (e.g., databases) to LLMs for improving performance and mitigating hallucinations.  
{\tech} belongs to a knowledge-intensive domain, where RAG can update security knowledge to guarantee the generated rule quality.  
This work only integrates prompt engineering (task decomposition, CoT, reflection, and few shots) into {\tech}  without RAG.
Note that prompt engineering and RAG are not mutually exclusive and can complement each other to improve LLMs' capabilities.

\textbf{Fine-Tuning} is to train a pre-trained LLM on a domain-specific dataset.
{\tech} can achieve a better performance when we tailor an LLM to the rule generation task.
Specifically, there are three requirements for fine-tuning LLM: (1) a pre-trained model (e.g., Llama 3.1), (2) a labeled domain dataset; and (3) training the LLM with Transformers. 
However, we lack a labeled domain dataset for fine-tuning LLM. 
The labeled data should be in the supervised format, denoted as \textit{(a malicious package, a rule)}.


\section{Related Work}
\label{sec:related}


\textbf{Software Supply Chain Attack}.
Nowadays, OSS ecosystems have millions of packages~\cite{constantinou2017empirical, ma2018constructing, ma2019world}, and the package dependencies are becoming very complex~\cite{serebrenik2015challenges, zimmermann2019small, zahan2021weak, pashchenko2020qualitative}.
Malicious packages~\cite{duan2020towards, guo2023empirical, Ohmssc2020, wy2022InstalltimeAtt, lidisa2022javabytecode, guo2023empirical} often contain code specifically crafted to perform unwanted or harmful actions on a system, including obfuscated code, data exfiltration, self-execution, typosquatting, dependency confusion, and backdoors. 
Pfretzschner and ben Othmane~\cite{pfretzschner2017identification} proposed a detection algorithm for dependency-based attacks on Node.js, and Staicu et al.~\cite{staicu2018understanding} proposed a deep understanding of the injection attack on Node.js.
Ladisa et al.~\cite{ladisa2023sok} proposed a general taxonomy of conceptual attack vectors in the OSS ecosystem.
Guo et al.~\cite{guo2023empirical} collected malicious packages in PyPI ecosystems, leveraged the case study to analyze the malicious behaviors.

Vulnerable packages~\cite{decan2018impact, decan2018evolution} refer to software libraries or components that contain security flaws, misconfigurations, or weaknesses that can be exploited by malicious actors.
Alfadel et al.~\cite{alfadel2021empirical} studied a collection of 550 vulnerabilities affecting 252 PyPi packages, and their analysis showed that vulnerabilities grew over time, and the most common was XSS vulnerabilities.
Ponta et al.~\cite{ponta2018beyond} proposed a code-centric scheme for detecting, analyzing, and mitigating vulnerabilities in software packages. 
Woo et al.~\cite{woo2021v0finder} traced a reported vulnerability back to its origin in the codebase, and combined static analysis, metadata extraction, and historical codebase tracking to identify the original vulnerability.

OSS ecosystems leak sensitive data and compromise users' privacy, caused by flaws, insecure configurations, and improper use of third-party dependencies.
Vaidya et al.~\cite{vaidya2019security} pointed out private information is leaked in the code of software packages, including key files and API keys embedded in the code.
Xiao et al.~\cite{xiao2021abusing} proposed an attack that abuses hidden attributes, which attackers can exploit to obtain confidential data, bypass security checks, and launch denial-of-service attacks.

\textbf{Large Language Model} is primarily based on the Transformer architecture with extensive pre-training on large-scale training data, and it has demonstrated remarkable advances across various domains. 
Previous works~\cite{wei2022chain, yao2023tree, yao2022react} proposed a planner to extend LLMs' capabilities for dealing with complex tasks, such as generalization and reasoning using task decomposition techniques such as Chain-of-thought~\cite{wei2022chain} and Tree-of-thought~\cite{yao2023tree}. 
Yao et al.~\cite{yao2022react} proposed a general strategy of self-reflection (called ReAct) based on the feedback of LLMs to improve their reasoning skills.
Similarly, Reflexion~\cite{shinn2023reflexion} and Chain of Hindsight~\cite{liu2023languages} use human feedback (e.g., errors) to fine-tune LLMs.
Another distinct approach is to use external resources to build an autonomous agent, enabling LLMs to interact with the environment (tools or APIs).
Wang et al.~\cite{wang2023voyager} proposed an embodied lifelong learning agent based on LLMs. 
Huang et al.~\cite{huang2022large} let an LLM self-improve its reasoning without supervised data by asking the LLM to lay out different possible results.

Meanwhile, LLMs can be applied in the software engineering domain, such as OpenAI Codex~\cite{open-codex} and GitHub Copilot~\cite{copilot}. 
Many prior works leveraged LLMs to resolve specific tasks in the software engineering domain~\cite{ma2023scope, sun2023automatic, pei2023can, xia2023keep, chen2021evaluating, fan2023large}.
Ma et al.~\cite{ma2023scope} and Sun et al.~\cite{sun2023automatic} explored the capabilities of LLMs when performing various program analysis tasks, including control flow graph construction, call graph analysis, and code summarization. 
There are many prior works using LLMs to resolve specific tasks in the security domain~\cite{chen2023teaching, ullah2023can}.
Pearce et al.~\cite{pearce2022pop} proposed to leverage LLM to help security professionals reverse engineer the binary application for automatically repairing vulnerabilities.
Li et al.~\cite{li2023hitchhiker} presented LLM capabilities in the static analysis of finding vulnerabilities. 
Feng and Chen~\cite{feng2023prompting} proposed to use LLM to replay Android bugs automatedly. Pearce et al.~\cite{pearce2023examining} examined zero-shot vulnerability repair using LLMs.





\textbf{Security Rules}. Security detection tools are widely used in the software engineering domain to detect security vulnerabilities, malicious packages, and privacy leaks.
YARA-scanner~\cite{naik2020evaluating, raff2020automatic, brengel2021yarix} is a tool for leveraging YARA rules to analyze the malicious features in the software package.
AppInspector~\cite{appInspector} includes 16 aspects and a total of 712 rules, which are used to check regular expression-based malicious patterns (e.g., reverse shell) in the text of each file in the package. 
Semgrep~\cite{semgrep} is a tool for pattern matching and searching in structured data, which can be used to detect sensitive information leakage, configuration errors, etc. 

\section{Conclusion}
\label{sec:conclusion}

In this paper, we demonstrate that LLMs can be an effective tool for generating rules in OSS ecosystems. 
{\tech} leverages LLMs to systematically analyze metadata and code, creating accurate YARA \& Semgrep rules without manual efforts, achieving a precision of 85.2\% and a recall of 91.8\%. 
Through its prototype implementation, {\tech} has proven its effectiveness and adaptability, showcasing its capability to enhance current security practices and outperform established tools.
Our work holds promise for future advancements in OSS security, particularly as a scalable solution to evolving SSC threats.

\section*{Acknowledgement}
We are grateful to our shepherd Xinda Wang and anonymous reviewers for their insightful feedback. 
The work was partially supported by the National Natural Science Foundation of China (No. 62272029 and No. 61972024).
\small
\bibliographystyle{IEEEtran}
\bibliography{ref/ref_software_supply_chain, ref/ref_online, ref/ref_algorithm, ref/ref_llm, ref/ref_ours}

\begin{thebibliography}{10}
\providecommand{\url}[1]{#1}
\csname url@samestyle\endcsname
\providecommand{\newblock}{\relax}
\providecommand{\bibinfo}[2]{#2}
\providecommand{\BIBentrySTDinterwordspacing}{\spaceskip=0pt\relax}
\providecommand{\BIBentryALTinterwordstretchfactor}{4}
\providecommand{\BIBentryALTinterwordspacing}{\spaceskip=\fontdimen2\font plus
\BIBentryALTinterwordstretchfactor\fontdimen3\font minus
  \fontdimen4\font\relax}
\providecommand{\BIBforeignlanguage}[2]{{%
\expandafter\ifx\csname l@#1\endcsname\relax
\typeout{** WARNING: IEEEtran.bst: No hyphenation pattern has been}%
\typeout{** loaded for the language `#1'. Using the pattern for}%
\typeout{** the default language instead.}%
\else
\language=\csname l@#1\endcsname
\fi
#2}}
\providecommand{\BIBdecl}{\relax}
\BIBdecl

\bibitem{oss_report}
Sonatype. (2021) {State of the software supply chain}.
  \url{https://www.sonatype.com/resources/state-of-the-software-supply-chain-2021}.

\bibitem{ssc-event}
E.Roth. (2021) {Open source developer corrupts widely-used libraries, affecting
  tons of projects.}
  \url{https://www.theverge.com/2022/1/9/22874949/developer-corrupts-open-source-libraries-projects-affected}.

\bibitem{pashchenko2020preliminary}
I.~Pashchenko, D.-L. Vu, and F.~Massacci, ``Preliminary findings on foss
  dependencies and security,'' 2020.

\bibitem{log4j}
C.~org. (2022) {Apache Log4j Vulnerability}.
  \url{https://www.cisa.gov/news-events/news/apache-log4j-vulnerability-guidance}.

\bibitem{semgrep}
semgrep org. (2019) {SemGrep rules for the security static analysis.}
  \url{https://github.com/semgrep/semgrep}.

\bibitem{naik2020evaluating}
N.~Naik, P.~Jenkins, R.~Cooke, J.~Gillett, and Y.~Jin, ``Evaluating
  automatically generated yara rules and enhancing their effectiveness,'' in
  \emph{2020 IEEE Symposium Series on Computational Intelligence (SSCI)}.\hskip
  1em plus 0.5em minus 0.4em\relax IEEE, 2020, pp. 1146--1153.

\bibitem{appInspector}
M.~org. (2023, accessible) {The tool identifies coding features of first or
  third party software components.}
  \url{https://github.com/microsoft/ApplicationInspector}.

\bibitem{ullah2023llms}
S.~Ullah, M.~Han, S.~Pujar, H.~Pearce, A.~Coskun, and G.~Stringhini, ``Llms
  cannot reliably identify and reason about security vulnerabilities (yet?): A
  comprehensive evaluation, framework, and benchmarks,'' \emph{arXiv preprint
  arXiv:2312.12575}, 2023.

\bibitem{pearce2023examining}
H.~Pearce, B.~Tan, B.~Ahmad, R.~Karri, and B.~Dolan-Gavitt, ``Examining
  zero-shot vulnerability repair with large language models,'' in \emph{2023
  IEEE Symposium on Security and Privacy (SP)}.\hskip 1em plus 0.5em minus
  0.4em\relax IEEE, 2023, pp. 2339--2356.

\bibitem{wang2024repository}
X.~Wang, R.~Hu, C.~Gao, X.-C. Wen, Y.~Chen, and Q.~Liao, ``A repository-level
  dataset for detecting, classifying and repairing software vulnerabilities,''
  \emph{arXiv preprint arXiv:2401.13169}, 2024.

\bibitem{li2023hitchhiker}
H.~Li, Y.~Hao, Y.~Zhai, and Z.~Qian, ``The hitchhiker's guide to program
  analysis: A journey with large language models,'' \emph{arXiv preprint
  arXiv:2308.00245}, 2023.

\bibitem{shang2024far}
X.~Shang, S.~Cheng, G.~Chen, Y.~Zhang, L.~Hu, X.~Yu, G.~Li, W.~Zhang, and
  N.~Yu, ``How far have we gone in stripped binary code understanding using
  large language models,'' \emph{arXiv preprint arXiv:2404.09836}, 2024.

\bibitem{pearce2022pop}
H.~Pearce, B.~Tan, P.~Krishnamurthy, F.~Khorrami, R.~Karri, and
  B.~Dolan-Gavitt, ``Pop quiz! can a large language model help with reverse
  engineering?'' \emph{arXiv preprint arXiv:2202.01142}, 2022.

\bibitem{guarddog}
E.~Wang. (2020) {The CLI tool that allows to identify malicious PyPI and npm
  packages}. \url{https://github.com/DataDog/guarddog}.

\bibitem{pkg-popular}
H.~van Kemenade. (2024) {Top PyPI Packages}.
  \url{https://hugovk.github.io/top-pypi-packages/}.

\bibitem{rule-code}
X.~Zhang, ``Malware detection rule generator.''
  \url{https://github.com/zhang-xr/RuleLLM}, 2024.

\bibitem{openai}
O.~Org. (2023) {The OpenAI API is used for a range of models and fine-tune
  custom models.} \url{https://platform.openai.com/docs/introduction}.

\bibitem{open-codex}
------. (2023) {OpenAI Codex: AI system that translates natural language to
  code.} \url{https://openai.com/blog/openai-codex}.

\bibitem{claude}
A.~Org. (2023) {Claude is a next generation AI assistant.}
  \url{https://claude.ai/}.

\bibitem{gemini}
G.~Org. (2023) {Google's Gemini family for the multi-modal model.}
  \url{https://poe.com/Gemini-Pro}.

\bibitem{wei2022chain}
J.~Wei, X.~Wang, D.~Schuurmans, M.~Bosma, F.~Xia, E.~Chi, Q.~V. Le, D.~Zhou
  \emph{et~al.}, ``Chain-of-thought prompting elicits reasoning in large
  language models,'' \emph{Advances in Neural Information Processing Systems},
  vol.~35, pp. 24\,824--24\,837, 2022.

\bibitem{yao2023tree}
S.~Yao, D.~Yu, J.~Zhao, I.~Shafran, T.~L. Griffiths, Y.~Cao, and K.~Narasimhan,
  ``Tree of thoughts: Deliberate problem solving with large language models,''
  \emph{arXiv preprint arXiv:2305.10601}, 2023.

\bibitem{shinn2023reflexion}
N.~Shinn, F.~Cassano, A.~Gopinath, K.~R. Narasimhan, and S.~Yao, ``Reflexion:
  Language agents with verbal reinforcement learning,'' in \emph{Thirty-seventh
  Conference on Neural Information Processing Systems}, 2023.

\bibitem{liu2023languages}
H.~Liu, C.~Sferrazza, and P.~Abbeel, ``Languages are rewards: Hindsight
  finetuning using human feedback,'' \emph{arXiv preprint arXiv:2302.02676},
  2023.

\bibitem{feng2020codebert}
Z.~Feng, D.~Guo, D.~Tang, N.~Duan, X.~Feng, M.~Gong, L.~Shou, B.~Qin, T.~Liu,
  D.~Jiang \emph{et~al.}, ``Codebert: A pre-trained model for programming and
  natural languages,'' \emph{arXiv preprint arXiv:2002.08155}, 2020.

\bibitem{numpy}
\BIBentryALTinterwordspacing
Numpy. (2009) numpy: Randomly permute a sequence. [Online]. Available:
  \url{http://docs.scipy.org/doc/numpy/reference/generated/numpy.random.permutation.html}
\BIBentrySTDinterwordspacing

\bibitem{scikit}
Scikit-learn. (2007) Machine learning library for the python language.
  \url{http://scikit-learn.org/stable/index.html}.

\bibitem{regex}
F.~Jeffrey. (2009) {Regular expression operations}.
  \url{https://docs.python.org/3/library/re.html}.

\bibitem{nltk}
NLTK. (2001) {A suite of libraries and programs for symbolic and statistical
  natural language processing.} \url{http://www.nltk.org/}.

\bibitem{langchain}
\BIBentryALTinterwordspacing
LangSmith. (2023) {LangChain, a unified platform for debugging, testing,
  evaluating, and monitoring your LLM applications.} [Online]. Available:
  \url{https://blog.langchain.dev/announcing-langsmith/}
\BIBentrySTDinterwordspacing

\bibitem{raff2020automatic}
E.~Raff, R.~Zak, G.~Lopez~Munoz, W.~Fleming, H.~S. Anderson, B.~Filar,
  C.~Nicholas, and J.~Holt, ``Automatic yara rule generation using
  biclustering,'' in \emph{Proceedings of the 13th ACM Workshop on Artificial
  Intelligence and Security}, 2020, pp. 71--82.

\bibitem{brengel2021yarix}
M.~Brengel and C.~Rossow, ``$\{$YARIX$\}$: Scalable $\{$YARA-based$\}$ malware
  intelligence,'' in \emph{30th USENIX Security Symposium (USENIX Security
  21)}, 2021, pp. 3541--3558.

\bibitem{llama}
M.~Org. (2023) {Llama 2: open source, free for research and commercial use.}
  \url{https://llama.meta.com/llama2/}.

\bibitem{vovk}
A.~Org. (2023) {Vovk — Advanced Yara rule generator}.
  \url{https://github.com/malienist/vovk?tab=readme-ov-file}.

\bibitem{xu2024benchmarkdatacontaminationlarge}
\BIBentryALTinterwordspacing
C.~Xu, S.~Guan, D.~Greene, and M.-T. Kechadi, ``Benchmark data contamination of
  large language models: A survey,'' 2024. [Online]. Available:
  \url{https://arxiv.org/abs/2406.04244}
\BIBentrySTDinterwordspacing

\bibitem{shi2023detecting}
W.~Shi, A.~Ajith, M.~Xia, Y.~Huang, D.~Liu, T.~Blevins, D.~Chen, and
  L.~Zettlemoyer, ``Detecting pretraining data from large language models,''
  \emph{arXiv preprint arXiv:2310.16789}, 2023.

\bibitem{constantinou2017empirical}
E.~Constantinou and T.~Mens, ``An empirical comparison of developer retention
  in the rubygems and npm software ecosystems,'' \emph{Innovations in Systems
  and Software Engineering}, vol.~13, no.~2, pp. 101--115, 2017.

\bibitem{ma2018constructing}
Y.~Ma, ``Constructing supply chains in open source software,'' in \emph{2018
  IEEE/ACM 40th International Conference on Software Engineering: Companion
  (ICSE-Companion)}.\hskip 1em plus 0.5em minus 0.4em\relax IEEE, 2018, pp.
  458--459.

\bibitem{ma2019world}
Y.~Ma, C.~Bogart, S.~Amreen, R.~Zaretzki, and A.~Mockus, ``World of code: an
  infrastructure for mining the universe of open source vcs data,'' in
  \emph{2019 IEEE/ACM 16th International Conference on Mining Software
  Repositories (MSR)}.\hskip 1em plus 0.5em minus 0.4em\relax IEEE, 2019, pp.
  143--154.

\bibitem{serebrenik2015challenges}
A.~Serebrenik and T.~Mens, ``Challenges in software ecosystems research,'' in
  \emph{Proceedings of the 2015 European Conference on Software Architecture
  Workshops}, 2015, pp. 1--6.

\bibitem{zimmermann2019small}
M.~Zimmermann, C.-A. Staicu, C.~Tenny, and M.~Pradel, ``Small world with high
  risks: A study of security threats in the npm ecosystem,'' in \emph{28th
  USENIX Security Symposium (USENIX Security 19)}, 2019, pp. 995--1010.

\bibitem{zahan2021weak}
N.~Zahan, L.~Williams, T.~Zimmermann, P.~Godefroid, B.~Murphy, and C.~Maddila,
  ``What are weak links in the npm supply chain?'' \emph{arXiv preprint
  arXiv:2112.10165}, 2021.

\bibitem{pashchenko2020qualitative}
I.~Pashchenko, D.-L. Vu, and F.~Massacci, ``A qualitative study of dependency
  management and its security implications,'' in \emph{Proceedings of the 2020
  ACM SIGSAC Conference on Computer and Communications Security}, 2020, pp.
  1513--1531.

\bibitem{duan2020towards}
R.~Duan, O.~Alrawi, R.~P. Kasturi, R.~Elder, B.~Saltaformaggio, and W.~Lee,
  ``Towards measuring supply chain attacks on package managers for interpreted
  languages,'' \emph{arXiv preprint arXiv:2002.01139}, 2020.

\bibitem{guo2023empirical}
W.~Guo, Z.~Xu, C.~Liu, C.~Huang, Y.~Fang, and Y.~Liu, ``An empirical study of
  malicious code in pypi ecosystem,'' in \emph{2023 38th IEEE/ACM International
  Conference on Automated Software Engineering (ASE)}.\hskip 1em plus 0.5em
  minus 0.4em\relax IEEE, 2023, pp. 166--177.

\bibitem{Ohmssc2020}
M.~Ohm, H.~Plate, A.~Sykosch, and M.~Meier, ``Backstabber's knife collection: A
  review of open source software supply chain attacks,'' in \emph{Detection of
  Intrusions and Malware, and Vulnerability Assessment}, C.~Maurice, L.~Bilge,
  G.~Stringhini, and N.~Neves, Eds.\hskip 1em plus 0.5em minus 0.4em\relax
  Cham: Springer International Publishing, 2020, pp. 23--43.

\bibitem{wy2022InstalltimeAtt}
\BIBentryALTinterwordspacing
E.~Wyss, A.~Wittman, D.~Davidson, and L.~De~Carli, ``Wolf at the door:
  Preventing install-time attacks in npm with latch,'' in \emph{Proceedings of
  the 2022 ACM on Asia Conference on Computer and Communications Security},
  ser. ASIA CCS '22.\hskip 1em plus 0.5em minus 0.4em\relax New York, NY, USA:
  Association for Computing Machinery, 2022, pp. 1139 -- 1153. [Online].
  Available: \url{https://doi.org/10.1145/3488932.3523262}
\BIBentrySTDinterwordspacing

\bibitem{lidisa2022javabytecode}
\BIBentryALTinterwordspacing
P.~Ladisa, H.~Plate, M.~Martinez, O.~Barais, and S.~E. Ponta, ``Towards the
  detection of malicious java packages,'' in \emph{Proceedings of the 2022 ACM
  Workshop on Software Supply Chain Offensive Research and Ecosystem Defenses},
  ser. SCORED'22.\hskip 1em plus 0.5em minus 0.4em\relax New York, NY, USA:
  Association for Computing Machinery, 2022, pp. 63 -- 72. [Online]. Available:
  \url{https://doi.org/10.1145/3560835.3564548}
\BIBentrySTDinterwordspacing

\bibitem{pfretzschner2017identification}
B.~Pfretzschner and L.~ben Othmane, ``Identification of dependency-based
  attacks on node. js,'' in \emph{Proceedings of the 12th International
  Conference on Availability, Reliability and Security}, 2017, pp. 1--6.

\bibitem{staicu2018understanding}
C.-A. Staicu, M.~Pradel, and B.~Livshits, ``Understanding and automatically
  preventing injection attacks on node. js,'' in \emph{Network and Distributed
  System Security Symposium (NDSS)}, 2018.

\bibitem{ladisa2023sok}
P.~Ladisa, H.~Plate, M.~Martinez, and O.~Barais, ``Sok: Taxonomy of attacks on
  open-source software supply chains,'' in \emph{2023 IEEE Symposium on
  Security and Privacy (SP)}.\hskip 1em plus 0.5em minus 0.4em\relax IEEE,
  2023, pp. 1509--1526.

\bibitem{decan2018impact}
A.~Decan, T.~Mens, and E.~Constantinou, ``On the impact of security
  vulnerabilities in the npm package dependency network,'' in \emph{Proceedings
  of the 15th international conference on mining software repositories}, 2018,
  pp. 181--191.

\bibitem{decan2018evolution}
------, ``On the evolution of technical lag in the npm package dependency
  network,'' in \emph{2018 IEEE International Conference on Software
  Maintenance and Evolution (ICSME)}.\hskip 1em plus 0.5em minus 0.4em\relax
  IEEE, 2018, pp. 404--414.

\bibitem{alfadel2021empirical}
M.~Alfadel, D.~E. Costa, and E.~Shihab, ``Empirical analysis of security
  vulnerabilities in python packages,'' in \emph{2021 IEEE International
  Conference on Software Analysis, Evolution and Reengineering (SANER)}.\hskip
  1em plus 0.5em minus 0.4em\relax IEEE, 2021, pp. 446--457.

\bibitem{ponta2018beyond}
S.~E. Ponta, H.~Plate, and A.~Sabetta, ``Beyond metadata: Code-centric and
  usage-based analysis of known vulnerabilities in open-source software,'' in
  \emph{2018 IEEE International Conference on Software Maintenance and
  Evolution (ICSME)}.\hskip 1em plus 0.5em minus 0.4em\relax IEEE, 2018, pp.
  449--460.

\bibitem{woo2021v0finder}
S.~Woo, D.~Lee, S.~Park, H.~Lee, and S.~Dietrich, ``$\{$V0Finder$\}$:
  Discovering the correct origin of publicly reported software
  vulnerabilities,'' in \emph{30th USENIX Security Symposium (USENIX Security
  21)}, 2021, pp. 3041--3058.

\bibitem{vaidya2019security}
R.~K. Vaidya, L.~De~Carli, D.~Davidson, and V.~Rastogi, ``Security issues in
  language-based sofware ecosystems,'' \emph{arXiv preprint arXiv:1903.02613},
  2019.

\bibitem{xiao2021abusing}
F.~Xiao, J.~Huang, Y.~Xiong, G.~Yang, H.~Hu, G.~Gu, and W.~Lee, ``Abusing
  hidden properties to attack the node. js ecosystem,'' in \emph{30th USENIX
  Security Symposium (USENIX Security 21)}, 2021, pp. 2951--2968.

\bibitem{yao2022react}
S.~Yao, J.~Zhao, D.~Yu, N.~Du, I.~Shafran, K.~Narasimhan, and Y.~Cao, ``React:
  Synergizing reasoning and acting in language models,'' \emph{arXiv preprint
  arXiv:2210.03629}, 2022.

\bibitem{wang2023voyager}
G.~Wang, Y.~Xie, Y.~Jiang, A.~Mandlekar, C.~Xiao, Y.~Zhu, L.~Fan, and
  A.~Anandkumar, ``Voyager: An open-ended embodied agent with large language
  models,'' \emph{arXiv preprint arXiv:2305.16291}, 2023.

\bibitem{huang2022large}
J.~Huang, S.~S. Gu, L.~Hou, Y.~Wu, X.~Wang, H.~Yu, and J.~Han, ``Large language
  models can self-improve,'' \emph{arXiv preprint arXiv:2210.11610}, 2022.

\bibitem{copilot}
M.~Org. (2023) {Copilot: The AI developer tool.}
  \url{https://github.com/features/copilot}.

\bibitem{ma2023scope}
W.~Ma, S.~Liu, W.~Wang, Q.~Hu, Y.~Liu, C.~Zhang, L.~Nie, and Y.~Liu, ``The
  scope of chatgpt in software engineering: A thorough investigation,''
  \emph{arXiv preprint arXiv:2305.12138}, 2023.

\bibitem{sun2023automatic}
W.~Sun, C.~Fang, Y.~You, Y.~Miao, Y.~Liu, Y.~Li, G.~Deng, S.~Huang, Y.~Chen,
  Q.~Zhang \emph{et~al.}, ``Automatic code summarization via chatgpt: How far
  are we?'' \emph{arXiv preprint arXiv:2305.12865}, 2023.

\bibitem{pei2023can}
K.~Pei, D.~Bieber, K.~Shi, C.~Sutton, and P.~Yin, ``Can large language models
  reason about program invariants?'' in \emph{International Conference on
  Machine Learning}.\hskip 1em plus 0.5em minus 0.4em\relax PMLR, 2023, pp.
  27\,496--27\,520.

\bibitem{xia2023keep}
C.~S. Xia and L.~Zhang, ``Keep the conversation going: Fixing 162 out of 337
  bugs for 0.42 each using chatgpt,'' \emph{arXiv preprint arXiv:2304.00385},
  2023.

\bibitem{chen2021evaluating}
M.~Chen, J.~Tworek, H.~Jun, Q.~Yuan, H.~P. d.~O. Pinto, J.~Kaplan, H.~Edwards,
  Y.~Burda, N.~Joseph, G.~Brockman \emph{et~al.}, ``Evaluating large language
  models trained on code,'' \emph{arXiv preprint arXiv:2107.03374}, 2021.

\bibitem{fan2023large}
A.~Fan, B.~Gokkaya, M.~Harman, M.~Lyubarskiy, S.~Sengupta, S.~Yoo, and J.~M.
  Zhang, ``Large language models for software engineering: Survey and open
  problems,'' \emph{arXiv preprint arXiv:2310.03533}, 2023.

\bibitem{chen2023teaching}
X.~Chen, M.~Lin, N.~Sch{\"a}rli, and D.~Zhou, ``Teaching large language models
  to self-debug,'' \emph{arXiv preprint arXiv:2304.05128}, 2023.

\bibitem{ullah2023can}
S.~Ullah, M.~Han, S.~Pujar, H.~Pearce, A.~Coskun, and G.~Stringhini, ``Can
  large language models identify and reason about security vulnerabilities? not
  yet,'' \emph{arXiv preprint arXiv:2312.12575}, 2023.

\bibitem{feng2023prompting}
S.~Feng and C.~Chen, ``Prompting is all your need: Automated android bug replay
  with large language models,'' \emph{arXiv preprint arXiv:2306.01987}, 2023.

\end{thebibliography}

\end{sloppypar}
\end{document}